# Experimental benchmark data for Monte Carlo simulated radiation effects of gold nanoparticles. Part I: Experiment and raw data analysis

**Hans Rabus[1], Philine Hepperle[1,2], Christoph Schlueter[3], Andrei Hloskovsky[3], Woon Yong Baek[1]**

[1] Physikalisch-Technische Bundesanstalt, Braunschweig and Berlin, Germany
[2] Leibniz Universität Hannover, Institute of Radioecology and Radiation Protection, Hannover, Germany
[3] Deutsches Elektronen-Synchrotron DESY, Hamburg, Germany

E-mail: hans.rabus@ptb.de

## Abstract

Electron emission spectra of gold nanoparticles (AuNPs) after photon interaction were measured over the energy range between 50 eV and 9500 eV to provide reference data for Monte Carlo radiation-transport simulations. Experiments were performed with the HAXPES spectrometer at the PETRA III high-brilliance beamline P22 at DESY (Hamburg, Germany) for photon energies below an above each of the gold L-edges, that is, at 11.9 keV, 12.0 keV, 13.7 keV, 13.8 keV, 14.3 keV, and 14.4 keV. The study focused on a sample with gold nanoparticles with an average diameter of 11.0 nm on a thin carbon foil. Additional measurements were performed on a sample with 5.3 nm gold nanoparticles and on reference samples of gold and carbon foils. Further measurements were made to calibrate the photon flux monitor, to characterize the transmission function of the electron spectrometer, and to determine the size of the photon beam. This allowed the determination of the absolute values of the spectral particle radiance of secondary electrons per incident photon flux. The paper presents the experimental and raw data analysis procedures, reviews the data obtained for the nanoparticle samples, and discusses their limitations.

## 1. Introduction

Gold nanoparticles (AuNPs) are considered potential radiosensitizers for radiation therapy [1–6]. AuNPs have been shown to increase the biological effectiveness of ionizing radiation in vitro and in vivo [5–9]. In these studies, the increase in average absorbed dose resulting from the higher absorption of radiation by the high-Z material gold (Au), as compared to biological matter or water, was much smaller than the observed effects. Therefore, these effects are often attributed to a local dose enhancement due to low-energy electrons from Auger cascades following core-shell ionizations of gold atoms [10–12]. Since this local enhancement is limited to microscopic dimensions in the range of a few 100 nm [10–15], it is often assessed by numerical simulations using radiation-transport Monte Carlo (MC) codes. The results from such MC simulations often show a wide range of outcomes between different studies [16,17]. However, a recent comparison exercise [18,19] demonstrated that some of the confusion in the literature is due to conceptional misunderstandings [13–15] or to a lack of quality assurance of simulation results [20,21].

Benchmarking MC simulations of the dosimetric effects of AuNPs with experimental data is challenging. Previous experimental studies were moreover limited to determining the macroscopic average dose increase [22–24]. In addition, the measurement of locally enhanced dose in the vicinity of AuNPs is not yet possible for two reasons. The first reason is that the range dose enhancement is limited to microscopic dimensions. And the second one is the low probability of photon interaction in an AuNP, which implies that only a small fraction of AuNPs contributes to the dose enhancement.

The expected number $\bar{n}_p$ of photon interactions in an AuNP is proportional to the absorbed dose $D$ and to the volume of the AuNP or the third power of its diameter, $d_g$ [13,20], and can be estimated by

$$\bar{n}_p = C_D \times d_g{}^3 \times D \tag{1}$$

The proportionality constant, $C_D$, depends on the photon energy spectrum. For the 50 kVp and 100 kVp X-ray spectra considered in above comparison exercise [18,19], $C_D$ is about $3.6 \times 10^{-7}$ Gy$^{-1}$ nm$^{-3}$ and $2.7 \times 10^{-7}$ Gy$^{-1}$ nm$^{-3}$, respectively.

Therefore, the expected number of photon interactions in a 13 nm AuNP is about $8 \times 10^{-4}$ and $6 \times 10^{-4}$, respectively, for a dose of 1 Gy. According to Dou et al. [25], diameters of about 13 nm may be optimal for clinical theranostic applications of AuNPs. Most radiobiological studies reviewed by Kuncic and Lacombe [6] used AuNP sizes well below 20 nm, that is, much smaller than the 50 nm and 100 nm considered in the MC comparison exercise.

The dosimetric effects of AuNPs result from electrons emitted after a photon interaction in the AuNPs. This is illustrated in Fig. 1, which shows the energy imparted by these electrons in radial water shells around the 50 nm and 100 nm AuNPs considered in the exercise. The energy imparted decreases rapidly within the first 150 nm from the AuNP. This is due to electrons with energies below about 3.5 keV being stopped, namely Auger electrons from the M and N-shells of Au, Coster-Kronig electrons from the L-shells of Au, and low-energy secondary electrons produced in the AuNP.

The plateau between 150 nm and 1000 nm is due to L-shell Auger electrons, for which a range in water between about 1.5 μm and 3 μm can be estimated from the ESTAR database [26]. The long-range energy deposition shown in the right panel of Fig. 1 is due to photoelectrons and Compton electrons leaving the AuNP.

The data shown in the left panel of Fig. 1 suggest that M-shell Auger electrons and the other low-energy electrons can contribute to the biological effects only when the AuNPs are located in the cell nucleus (which is generally not observed [27]) or near the nucleus' membrane in the cytoplasm. In contrast, AuNPs from a much larger volume of the cytoplasm

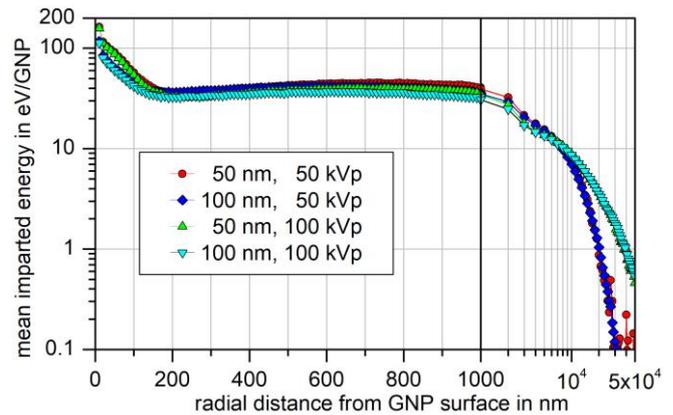

**Fig. 1:** Mean energy imparted in 10 nm thick spherical shells around an AuNP undergoing a photon interaction for two AuNP sizes and photon energy spectra. (Figure reproduced unchanged under CC-4.0-BY license from Rabus et al., Intercomparison of Monte Carlo calculated dose enhancement ratios for gold nanoparticles irradiated by X-rays: Assessing the uncertainty and correct methodology for extended beams, Physica Medica 84 (2021) 241–253. [15]) The data are from a multicenter study and represent the mean value of the results (from different participants) that passed the plausibility tests. The solid vertical line indicates a change from a linear to a logarithmic x-axis. In the right panel, the data points indicate the average over the 1 μm thick spherical shells used in the simulations for radial distances exceeding 1000 nm.

can contribute to the biological effects via emitted L-shell Auger electrons.

In the MC comparison exercise, the largest differences between simulations performed with different codes were seen in the ancillary results for the energy spectra of emitted electrons [18,19], especially at low energies. Discrepancies remained for some results after all necessary corrections were made to the originally reported ones [20]. A major reason for these remaining discrepancies was that one code used a newly developed dataset of cross sections for electron interactions in gold at energies down to the ionization threshold, which was derived from theoretical models [28]. Most other codes either used empirical extrapolations of the cross-section data from the Livermore database to lower energies or used the energy cut-offs implied by the energy ranges of that database. Recently, cross-sections for electron transport at low energies in gold have also become available in the Geant4-DNA code system [29,30].

The present project was initiated with the goal of providing an experimental dataset of electron emissions from gold nanoparticles as a benchmark for MC simulations, and thus indirectly, for the electron cross-section datasets implemented in the MC codes. Due to the relevance described above, the focus was placed on the energy range of Au L-shell Auger electrons and below.

Electron spectrometers require the samples under investigation to be in an ultra-high vacuum environment. Electron spectra emitted from AuNPs in vacuum should be very similar to those emitted from AuNPs in water, the only difference being that electrons may be emitted twice after backscattering in the water surrounding the AuNP. This is expected to be a negligible effect.

The low photon interaction probability in AuNPs makes such experiments challenging as high photon flux rates are required to achieve a reasonable signal-to-noise ratio. Therefore, the experiments were conducted in a synchrotron radiation facility. The beams used had photon fluence rates between $1\times10^{16}$ cm$^{-2}$ s$^{-1}$ and $5\times10^{16}$ cm$^{-2}$ s$^{-1}$, corresponding to absorbed dose rates to water between 230 kGy/s and 40 kGy/s. To further optimize the signal-to-noise ratio and to investigate the contributions of the Auger cascades from the different Au L-shells, the photon energies were chosen slightly above and slightly below each of the corresponding binding energies.

Smaller sizes of roghly 5 nm and 10 nm were used instead of AuNPs with the sizes considered in the code comparison exercise. These sizes of nanoparticles were used in the studies of Kim et al. [31] and Hainfeld et al. [32] and also in ongoing research by collaborating radiobiology groups at the radiooncology department of the University Medical Center Hamburg-Eppendorf (Hamburg, Germany) and Université de Namur (Namur, Belgium), which provided the AuNPs used in this study. According to Dou et al. [25], these AuNP sizes appear to be more clinically relevant than smaller or larger particles.

In this first part of the paper, the experimental procedures and data analysis methodology are described, with the emphasis on obtaining absolute results. The second part of the paper deals with the comparison of the measured data for a thin gold foil with MC simulations [33]. In the third part of the paper, the measured data will be used for benchmarking the "radial" code [34]. The detailed line shape analysis of the results on AuNP samples and a comparison with simulations will be presented in the fourth part.

This work contains three supplements describing auxiliary experiments and data analyses to characterize the experimental setup, which allowed the determination of results on an absolute scale. A fourth supplement shows additional results from measurements on the samples studied. The figures and tables in these supplements are referred to with the suffix "Sx-" before the figure or table number, where "x" is the supplement number.

## 2. Materials and Methods

### 2.1 Experimental setup

The measurements were performed at the high-brilliance beamline P22 of PETRA III at DESY (Hamburg, Germany), which offers photon energies between 2.4 keV and 15 keV at a photon flux in the order of $10^{13}$ s$^{-1}$ and is equipped with a high-resolution hard X-ray photoelectron spectroscopy (HAXPES) instrument that is capable of measuring electron energies up to 10 keV [35]. The undulator beam is monochromatized by an LN$_2$-cooled double crystal monochromator (DCM). For the present experiments, the Si(111) crystals of the monochromator were used, and the undulator was operated at its third harmonic.

The HAXPES electron spectrometer is a SPECS Phoibos 225 HV hemispherical analyzer (SPECS Surface Nano Analysis Ltd., Germany) mounted on an ultrahigh vacuum analysis chamber (base pressure $\sim 2 \times 10^{-9}$ mbar) such that its optical axis is horizontal and at right angles to the photon beam. The analysis chamber was equipped with a fully motorized 5-axis manipulator with three translational and two rotational degrees of freedom [35] and an attached load-lock chamber (average base pressure of $2 \cdot 10^{-9}$ mbar). Samples mounted on standard wedge-shaped copper sample holders manufactured at DESY (DESY, Germany) were introduced into the load-lock chamber and stored on a sample carousel. A long arm transfer allowed samples to be exchanged between the sample carousel and the manipulator in the analysis chamber.

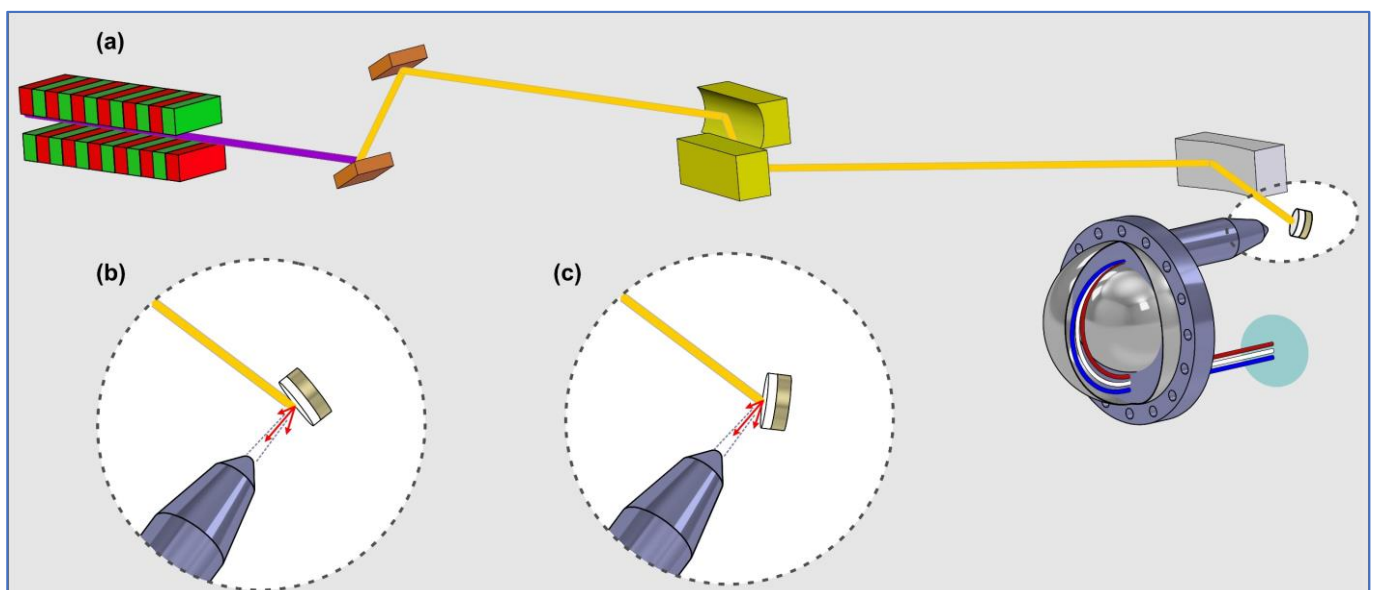

Fig. 2: (a) Schematic representation of the DESY P22 beamline with the HAXPES spectrometer. The leftmost element is the undulator, which emits a beam of broadband synchrotron radiation (blue line) which is spectrally filtered by the double-crystal monochromator. The resulting narrowband photon beam (orange line) is then focused with the mirrors (color) to the measurement position on the sample. The emitted electrons are detected with the hemispherical mirror analyzer. (b) and (c) show close-up images of the region around the sample indicated by the dashed line in (a). The photon beam (orange line) hits the sample surface at a grazing angle of incidence of (b) 15° and (c) 60°. The electrons emitted from the sample (arrows) are detected within the acceptance angle (boudaries indicated by dashed lines) of the spectrometer lens.

To monitor the photon flux, the current of a photodiode detecting emitted electrons and scattered photons from a thin carbon foil that was placed in the beamline in front of the last focusing mirror was recorded with one of the channels of a four-channel digitizer.

## 2.2 Samples and sample preparation

The main samples to be investigated consisted of citrate-stabilized gold nanoparticles (AuNPs) coated with polyethylene glycole (PEG) 11-mercaptoundecanoic acid (MUA). They were manufactured by the University of Hamburg (Hamburg, Germany) and were provided by the Medical Center Hamburg-Eppendorf (UKE, Hamburg, Germany). Their size distribution had a mean diameter of 11 nm and a standard deviation of 0.9 nm [36]. (Note that the value of 1.8 nm given in [36] is two standard deviations.) The AuNP samples for the experiments were prepared on a 50 nm thick self-supporting carbon foil (specified mass per area 10 µg/cm², purity 99.997 %) supplied by Goodfellow (Hamburg, Germany) on temporary glass supports.

The carbon foil was suspended over the 5 mm holes of a 1 mm thick aluminum support of rectangular shape and dimensions 22 mm × 13 mm, manufactured in a mechanical workshop of PTB (Braunschweig, Germany). The AuNP solution was stirred to homogenize the solution. A 0.5 µL drop of AuNP solution was cast onto the carbon microsheet in the center of one of the two holes in the sample holder using an

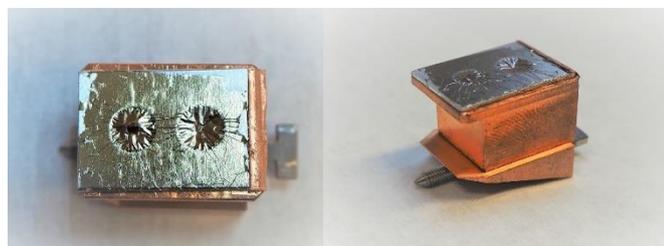

Fig. 3: Photographs of an aluminum support covered with a 50 nm thick self-supporting carbon foil mounted on a wedge-shaped DESY copper sample holder. In the top view on the left, the dark spot in the center of the left aperture is an AuNP sample deposited by the drop-casting method. The part of the carbon foil covering the right aperture was used for reference measurements.

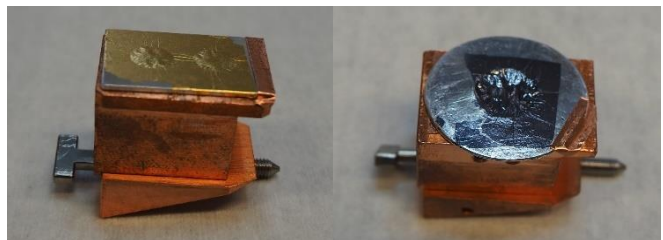

Fig. 4: Photographs of the other samples used in the measurements mounted on wedge-shaped copper DESY sample holders. Left: A 100 nm thick gold foil attached to an aluminum support with two 5 mm apertures. Right: The 5.3 nm AuNP sample provided by Université de Namur.

Eppendorf® pipette. The drop was allowed to dry for 24 hours at room temperature under ambient air.

AuNP samples prepared on carbon microsheets following the same protocol were examined by X-ray photoelectron spectroscopy (XPS) and scanning transmission electron microscope (STEM) to determine the size, shape, and surface coverage of the gold nanoparticles [36,37]. The sample used in the measurements was examined by STEM after the beamtime.

For the experiments, the aluminum supports of the samples were attached with double-sided adhesive copper tape to standard wedge-shaped copper sample holders used at DESY (Fig. 3).

Two other samples were also used in the experiments (Fig. 4). One consisted of a 100 nm thick gold foil mounted on an aluminum support of the same type. The second sample consisted of nanoparticles with an average diameter of 5.3 nm coated with PEG. They were deposited by ion beam sputtering on a 100 nm thick carbon foil suspended on a cylindrical aluminum support with an outer diameter of 20 mm and a 5 mm aperture in the center. This sample was fabricated at the University of Namur (Namur, Belgium).

## 2.3 Performed measurements of electron spectra

Electron emission spectra of AuNP and reference samples were recorded for electron energies between 50 eV and 9.5 keV in 1 eV steps. The measurements were performed for six photon energies corresponding to values slightly below and above the gold L-edges (11.9 keV, 12.0 keV, 13.7 keV, 13.8 keV, 14.3 keV, 14.4 keV). To facilitate the correction of its energy-dependent transmission, the spectrometer was operated in a fixed retarding ratio mode. Different values for the retarding ratio were used in different energy ranges (Table 1).

For the nanoparticle samples, data were recorded for two different grazing-incidence angles, namely 15° and 60°. A small grazing-incidence angle leads to a larger irradiated area and thus to a better signal-to-noise ratio. At the same time, the sensitivity to lateral movements of the photon beam increases. A 15° grazing-incidence angle is the commonly used optimal choice for a trade-off between the two effects.

At a small grazing angle of incidence, the scattering of emitted electrons by neighboring AuNPs in the surface plane is reduced. This scattering is best studied at normal incidence of the radiation, but then the surface seen by the spectrometer would disappear. An incidence angle of 60° is a good compromise between studying in-plane scattering and obtaining a reasonable signal.

For the gold and carbon foil samples, measurements were only taken at 15° grazing incidence, and for the carbon foil, only for the highest photon energy.

During the measurements, the count rate of the HAXPES and the current signal of the photon flux monitor were recorded simultaneously for each electron energy data point. To prevent the HAXPES electron detectors from overloading, the photon flux was reduced for some samples and energy ranges by moving attenuators that were installed upstream of the photon flux monitor into the beam.

Dark current suppression was repeatedly performed on the CAEN multi-channel electrometer used to measure the photon flux monitor signal. The remaining dark current was on the order of 10 pA and is treated as an uncertainty. The measured currents were on the order of several tens of nA (see Table S1.1 in Supplement 1).

## 2.4 Data analysis

The data analysis was based on the following measurement equation for the spectral particle radiance per photon flux, $d^3\varepsilon/dA\,dE\,d\Omega$ of electrons emitted from the irradiated sample. When all electrons passing the energy analyzer are registered, $d^3\varepsilon/dA\,dE\,d\Omega$ is the number of emitted electrons per incident photon, per electron energy interval, per solid angle, and per surface area (averaged over the irradiated area):

$$\frac{d^3\varepsilon(E_e)}{dA\,dE\,d\Omega} = \frac{\dot{N}_m}{I_0 E_e} G(R_r,R) \frac{R}{R_r} \frac{Q_0(E_p) E_r}{T(E_{ar},R_r)} \frac{A_r}{A_b} \sin\theta \quad (2)$$

In Eq. (2), $\dot{N}_m$ is the measured count rate at the set energy $E_e$ and the retarding ratio $R$ of the electron spectrometer, $I_0$ is the measured current of the photon flux monitor, and $Q_0$ is the average detected charge per photon passing the photon flux monitor. $E_r$ and $R_r$ are reference values for the kinetic energy and the retarding ratio for which the spectrometer transmission $T$ is known at the pass energy $E_{ar} = E_r/R_r$. $A_r$ is the beam cross-sectional area used in electron ray tracing for determining the transmission, $A_b$ is the actual beam cross-sectional area, and $\theta$ is the angle of grazing incidence of the photon beam on the sample surface. The factor $\sin\theta$ takes into account that the surface area hit by the photon beam is increased by a factor of $1/\sin\theta$. The quantity $G$ in Eq. (2) is the ratio of the spectrometer transmission at a reference pass energy $E_{ar}$ and retarding ratio $R_r$ to the one for retarding ratio $R$. Details on the determination of the photon flux, photon

Table 1: Energy ranges and retarding ratios used in the electron spectra measurements. The energy step was always 1 eV. The retarding ratio for the first energy range was generally $R = 1$, except for the measurements of the gold foil, where $R = 10$ was used.

| Energy range | Start energy / eV | End energy / eV | Retarding ratio $R$ |
|---|---|---|---|
| 1 | 50 | 120 | 1 (10) |
| 2 | 100 | 1200 | 10 |
| 3 | 1000 | 3500 | 50 |
| 4 | 3500 | 6500 | 50 |
| 5 | 6000 | 9500 | 100 |

beam size, and spectrometer transmission function are given in Supplements 1, 2, and 3.

In a measurement run on a sample, five datasets of emitted electron spectra were recorded with the HAXPES spectrometer, using different retarding ratios $R$ depending on the scanned energy ranges (Table 1). To obtain a joint dataset, the individual datasets were first normalized to the photon flux monitor signal. This was done to remove the temporal variation of the photon flux. (It did not, however, remove the negligibly small glitches at storage ring top-ups due to differences between the HAXPES dwell time and the integration time of the CAEN used for measurement of the photon flux monitor current.)

The data were then divided by the electron energy to compensate for the proportionality with $E$ of the spectrometer transmission in the fixed retarding ratio mode. In the next step, for each pair of adjacent energy scans with retarding ratios $R_1$ and $R_2$, the ratios of the two measurements were calculated for all common energy points in the overlap region. Mean and standard deviation were used as an estimate and the uncertainty of the ratio $G(R_1,R_2)$. Appropriate combinations of these ratios were then used to calculate the $G(R_r,R)$ ratios in Eq. (2) for each scan range. After multiplying the data in each scan range by the respective factor $G(R_r,R)$, the datasets were merged. In the overlap intervals between successive energy ranges, the average values of the two scans were used. Finally, the merged dataset was multiplied by the overall calibration factor, this being the fraction given as last factor in Eq. (2), to obtain the spectral particle radiance of electrons per incident photon flux.

# 3. Results

## 3.1 Overview

To get a first impression of the measurement results, Fig. 5 shows the obtained electron energy spectra of the 11 nm AuNP samples for the two photon energies 11.9 keV and 14.4 keV. The first energy is slightly below the Au L₃ binding

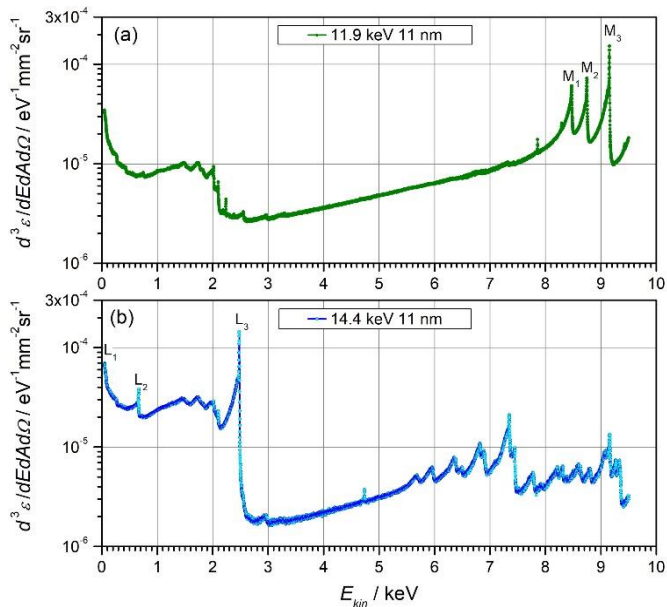

Fig. 5: Measured electron energy spectra for irradiation of the sample with 11 nm gold nanoparticles on a 50 nm carbon foil with photons of energies (a) 11.9 keV and (b) 14.4 keV, i.e. below the Au $L_3$ and above the Au $L_1$ absorption edges, at 15° grazing-incidence angle.

energy, so only electrons in the M and outer shells can be excited by photoabsorption.

The spectrum in Fig. 5(a) is dominated by the photoabsorption peaks of the $M_1$, $M_2$, and $M_3$-shells, which exhibit pronounced low-energy tailing. They are overlaid by a background consisting of the low-energy tailing of the $M_4$ and $M_5$ photoelectron lines, which peak at energies beyond the range studied here. In the energy range between 3 keV and 8 keV, this spectrum is generally characterized by the absence of spectral features, except for a sharp peak at about 7850 eV, which is the K-shell photoabsorption peak of potassium.

In the energy range between 0.8 keV and 2 keV, the spectrum contains the Auger electrons produced in non-radiative filling of vacancies in the M-shells. The large number of different transitions with energies close to each other results in a quasi-continuum, which has also been reported by other authors [38].

In the electron emission spectrum produced by 14.4 keV photons (blue symbols), a large number of lines can be seen in the energy range between 5.5 keV and 9.5 keV. Since the M-shell photoelectrons have shifted to higher energies outside the range covered by the measurements, these lines are due to Auger transitions when vacancies in the L-shells are filled. At this photon energy, all L-shells of gold can be excited, leading to the intense photoelectron peaks at about 0.8 keV ($L_2$) and 2.5 keV ($L_3$). (The $L_1$ line coincides with the low-energy secondary electron peak close to the lowest detected electron energy.)

These peaks and their tails strongly overlap with the M-shell Auger electrons, highlighting the complexity of a quantitative line shape analysis in this energy range. Changes in the intensity of the M Auger lines are expected for the higher photon energy due to the enhanced production of M-shell Auger electrons in the de-excitation cascade following photoabsorption in the L-shells of gold. In addition, the Coster-Kronig electrons produced when an L-shell vacancy is filled by another electron from a higher L-shell are also expected to appear in this energy range [39].

It should be noted that the sharp peak at about 4.7 keV appearing at about 2.2 keV in the spectrum for 11.9 keV photons corresponds to the K-shell photoelectron of zinc. Like the potassium peak mentioned earlier, this suggests surface contamination from residues of the chemicals used in processing the gold nanoparticles during sample preparation. It is also worth noting that there is no evidence of a K-shell photopeak of Cu in the samples (which would appear at about 3 keV and 6.5 keV for 11.9 keV and 14.4 keV photons, respectively.)

### 3.2 Photoelectron spectra

To further characterize the chemical composition of the AuNP and gold foil samples, Fig. 6 shows the emitted electron spectra produced by (a) 11.9 keV and (b) 14.4 keV photons as a function of binding energy in the energy ranges between 2.4 keV and 4.1 keV and between 11.6 keV and 14.4 keV, respectively. The vertical dashed lines show the binding energies of the atomic levels according to the evaluated atomic data library (EADL) [39,40]. The element is coded by symbol type and color, and the atomic shell is indicated as a label above each symbol.

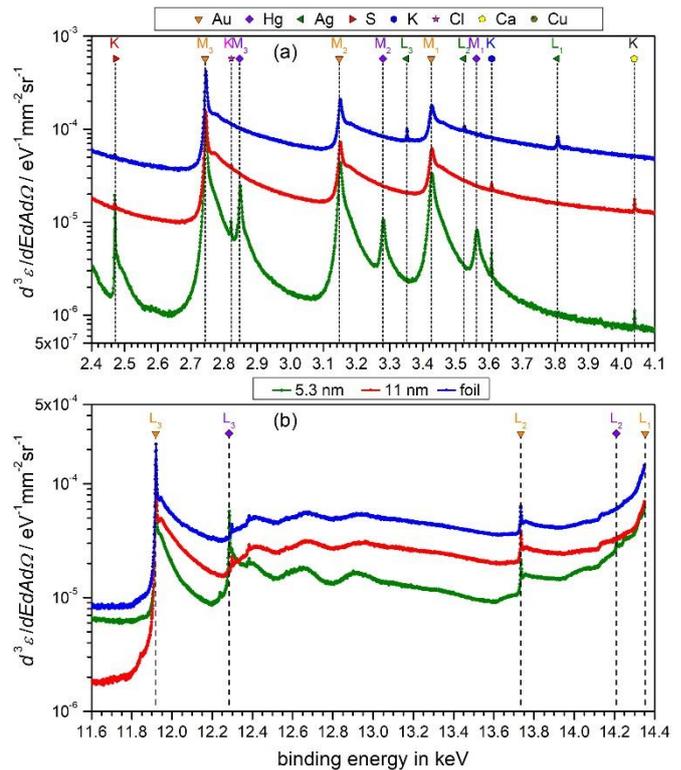

Fig. 6: Electron spectra measured with incident photons of energy (a) 11.9 keV (i.e., slightly below the Au $L_3$ edge) and (b) 14.4 keV (i.e. slightly above the Au $L_1$ edge). The vertical dashed lines indicate the binding energy values from the EADL for the elements and atomic shells indicated by the symbols and labels at the top of the diagram panel. The blue curves correspond to the 100 nm gold foil, the red curves to the sample with 11 nm gold nanoparticles on a 50 nm carbon foil, and the green curves to the sample with 5.3 nm gold nanoparticles on a 100 nm carbon foil.

In addition to the photoelectron lines from the M-shells of gold, lines corresponding to several other elements can be identified in Fig. 6(a). The photoelectron peak from the K-shell of sulfur is expected for the nanoparticle samples due to the presence of a thiol group that binds the polymer passivation layer to the nanoparticle surface. On the logarithmic scale it appears more pronounced for the smaller nanoparticles, which have a higher proportion of surface atoms. The presence of photoelectron peaks of calcium, chlorine, and potassium is due to the chemicals used in the treatment of the samples. The gold foil shows significant contamination with silver and copper (Fig. 7(a)), which is due to the good alloyability of the three elements.

The observation of a significant signal of mercury (Hg) photoelectrons in the spectra of the 5.3 nm gold nanoparticles was a surprise. This contamination is attributed to potential impurities of the gold substrate used in the dry physical production process. Gold and mercury are also easily alloyed. Using the peak height with respect to the nearest minimum at higher energies and the photo-absorption cross-sections for the $M_3$, $M_4$, and $M_5$-shells of Au and Hg, the molar ratio of Hg to Au was estimated to be about 1:6.3. (The cross-section values were obtained by fitting a power law to the data from the evaluated photon data library [41] for photon energies between 9.0 and 12.5 keV.) Thus, when the Hg atoms are inside the nanoparticles, their mole fraction amounts to $(13.7 \pm 0.5)$ %. If the Hg atoms are not inside the nanoparticle, but are bound to its surface, about every second surface atom in the AuNP has an Hg atom as a neighbor. (At a diameter of 5.3 nm, about 30 % of the gold atoms are located at the AuNP surface.)

Fig. 6(b) shows the energy spectra of the three samples for 14.4 keV photons in the energy range of the binding energies of the gold L-shells. Again, the dashed lines indicate the binding energies according to the EADL. It can be seen that for both, the gold and mercury lines, the peak positions coincide with the EADL value within ±1 eV. This is taken as evidence that the electron and photon energy scales are also correct within these limits.

Fig. 7 illustrates the dependence of the Cu-K photoelectron peak in the gold foil measurements on photon energy. The peaks shown in Fig. 7(a) were fitted with a Lorentz profile to obtain the peak area. The ratio of this peak area to the Cu K-shell photoabsorption cross-section is shown in Fig. 7(b). The latter was obtained by extracting data for energy points between 10 keV and 15 keV from the database provided with the Penelope code and interpolating based on a power law fit to these data. The relative size of the error bars in Fig. 7(b) is equal to the relative uncertainty of the peak area resulting from the fit to the data.

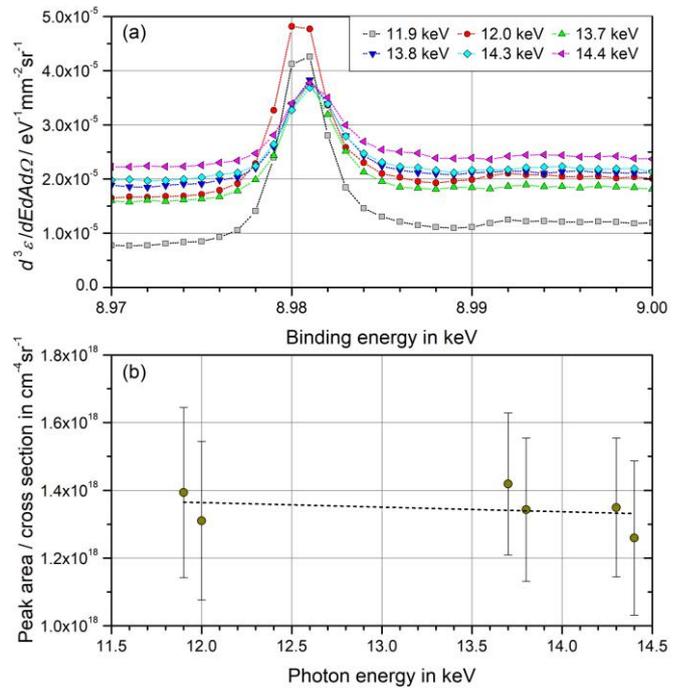

Fig. 7: (a) Spectral particle radiance per photon flux from the Au foil in the binding energy range of the K-shell of Cu for the measurements at different photon energies. (b) Photon energy dependence of the ratio of the photoelectron peak area to the photoabsorption cross-section of the K-shell of Cu.

The fact that the ratio of the peak area to the cross-section is essentially constant demonstrates the consistency of the data analysis procedure and that the energy dependence of the photon beam monitor (Supplement 1) and of the electron spectrometer transmission (Supplement 2) have been correctly evaluated.

### 3.3 Au L Auger electron energies

A completely different situation arises when considering the LMM Auger electrons of gold. The electron spectra measured for the three gold samples after excitation with photons of 11.9 keV and 12.0 keV are shown in Fig. 8 in the kinetic energy range between 5 keV and 8 keV. This region is the range of gold $L_3MM$ Auger electrons according to the EADL [39]. The respective electron emission peaks are expected for the 12.0 keV photon energy, whereas the $L_3$-edge of gold cannot be excited by 11.9 keV photons. The small features seen in the lowest curve of the 11.9 keV measurement on the 5.3 nm AuNP samples suggest that the incident photon beam may contain a small fraction of higher-energy photons from the higher orders of the monochromator and undulator. In contrast to the data from the 11 nm AuNPs and the gold foil, the 11.9 keV measurements of this sample do not show the monotonous increase with electron energy. This is due to a background of electrons produced in the sample holder that will be discussed in more detail in the next sections.

The peaks were assigned in Fig. 8 based on ascending energy and the peak intensities given in the EADL [39]. It is worth noting that the peaks appear at the same energies for all samples, although for the larger AuNPs and the Au foil, the low-energy tailing is very pronounced. This suggests that any effects of the local environment of the absorbing atoms in the nanoparticles are already fully active for the 5.3 nm AuNPs. It should be noted that these nanoparticles contain about 4000 atoms.

The energies of the peak maxima are listed in Table 2 together with the corresponding values from the EADL [39]. Significant discrepancies are observed for all peak energies, in some cases exceeding 100 eV. These shifts to lower energies are expected since the Auger energies reported in the EADL are calculated from differences of the binding energies of the electron levels involved. It is not taken into account that the de-excitation occurs for a positive ion and that there may be a rearrangement of the orbitals and their binding energies before the non-radiative transition occurs. This warrants a separate investigation, which will be presented in the third part of the paper, where the experimental data are used to benchmark the "radial" code [34].

### 3.4 Electron energy spectra of the 11 nm AuNP sample

Measurements were performed on the 11 nm AuNP samples at 15° and 60° grazing incidence of the photons. The rationale was that the two different irradiation geometries would allow discriminating effects of multilayers (which are more pronounced for 60° grazing incidence) and monolayers (at 15°), as well as the effect of impurities, which have been reported in other work [42–46]. In the absence of multilayers and impurities, the energy emission spectra for the two incidence angles should match if the nanoparticle surface density is the same in both measurements.

Table 2: Energy values of the Auger lines identified in Fig. 8 compared to the values in the EADL.

| Assignment | Experiment $E_{peak}$ / eV | EADL $E_{kin}$ / eV | Prob. | Difference in eV |
|---|---|---|---|---|
| $L_3M_1M_1$ | - | 5119.2 | 0.08 % | - |
| $L_3M_1M_2$ | - | 5376.3 | 0.04 % | - |
| $L_3M_2M_2$ | - | 5633.4 | 0.01 % | - |
| $L_3M_1M_3$ | 5684.3 | 5786.6 | 2.18 % | -102.3 |
| ? | 5715.5 | - | - | - |
| $L_3M_2M_3$ | 5963.6 | 6043.7 | 3.66 % | -80.1 |
| $L_3M_1M_4$ | 6139.8 | 6219.9 | 0.26 % | -80.1 |
| $L_3M_1M_5$ | 6228.0 | 6308.4 | 0.42 % | -80.4 |
| $L_3M_2M_4$ | 6356.2 | 6477.0 | 0.18 % | -120.8 |
| $L_3M_3M_3$ | 6380.1 | 6454.0 | 4.68 % | -63.9 |
| $L_3M_2M_5$ | 6501.9 | 6565.5 | 1.55 % | -63.6 |
| $L_3M_3M_4$ | 6822.2 | 6887.3 | 4.90 % | -65.1 |
| $L_3M_3M_5$ | 6896.0 | 6975.8 | 6.72 % | -79.8 |
| $L_3M_4M_4$ | 7277.7 | 7320.6 | 0.68 % | -42.9 |
| ? | 7317.2 | - | - | - |
| $L_3M_4M_5$ | 7348.9 | 7409.1 | 12.8 % | -60.2 |
| ? | 7415.5 | - | - | - |
| $L_3M_5M_5$ | 7449.8 | 7497.6 | 8.56 % | -47.8 |

It was found during the beamtime (cf. Fig. S4.1) and confirmed by the STEM measurements performed afterwards [36] that the surface is non-uniformly covered with AuNPs. Therefore, different measurement runs probing different regions of the sample resulted in different values of particle radiance. This is illustrated in Fig. 9, where results from measurements at two energies and 15° grazing incidence of the photon beam are plotted.

Fig. 9(a) shows a comparison of two measurements at 12 keV photon energy performed on almost the same surface area of the sample. In this case, there is good agreement of the two spectra with some notable deviations at smaller electron energies. This is explained by the fact that in the first experiment the energy scan over the second energy range (100 eV - 1.2 keV) initially failed and was only repeated after a longer interval (about 6 hours) at a slightly different spatial position on the sample.

Fig. 9(b) shows three measurements at 14.4 keV photon energy in different locations on the sample. Measurement position 1 was a spot that gave a maximum signal and was probed first to obtain a good signal-to-noise ratio (cf. Fig. S4.1(a) and (b) in Supplement 4). The posterior STEM investigations of the sample indicated the occurrence of so-called "coffee rings" on the sample where the AuNPs occur as multilayers [36]. Therefore, position 1 is attributed to a region on the sample surface with multilayer coverage. Position 5 was also located in the outer parts of the sample (Fig. S4.1 (c) and (d)) but is expected to be an area where monolayer coverage prevails. Position 6 was located in the central region of the sample (Fig. S4.1 (e) and (f)) where the AuNPs occur in (sub-)monolayers [36].

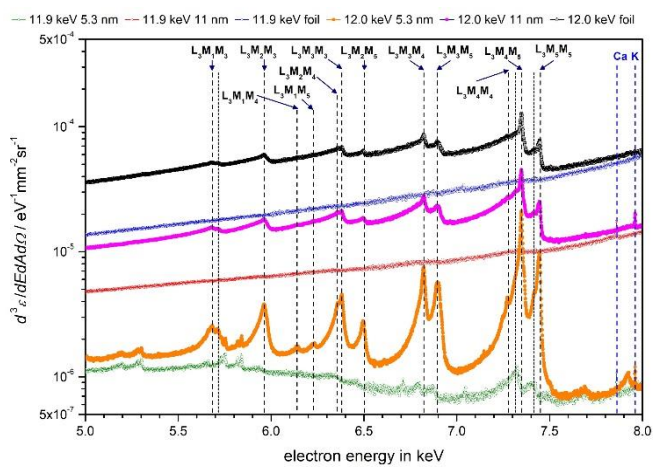

Fig. 8: Electron emission spectra in the kinetic energy range of the gold $L_3MM$ Auger electrons measured with photon excitation energies slightly below and above the Au $L_3$ edge (11.9 keV and 12.0 keV). The vertical dashed lines indicate the energy values of the identified gold Auger transitions (as well as the kinetic energies of Ca K-shell photoelectrons). The curves at the top are from a 100 nm gold foil, the curves in the middle correspond to measurements on 11 nm gold nanoparticles on a 50 nm carbon foil, and the curves at the bottom are for 5.3 nm gold nanoparticles on a 100 nm carbon foil.

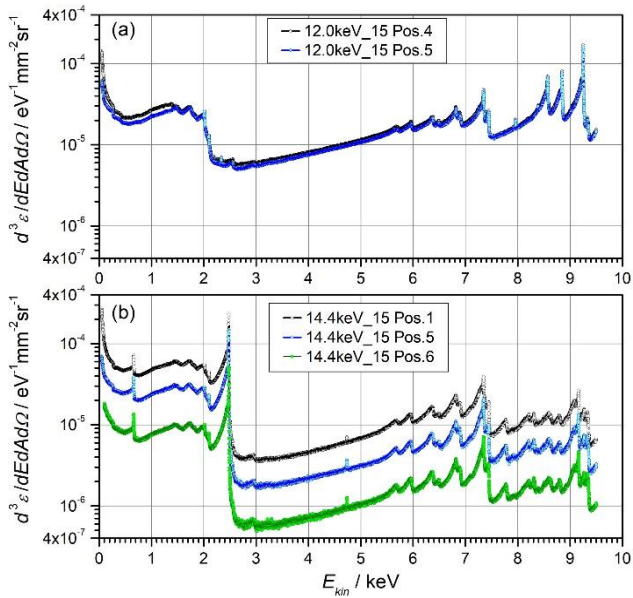

Fig. 9: Results for the spectral particle radiance per photon flux of the 11 nm AuNP sample measured at different spots on the 11 nm AuNP sample measured with photon energies of (a) 12.0 keV and (b) 14.4 keV.

As can be seen in Fig. 9(b), the spectra at different coverages appear to have the same relative energy dependence with an almost constant offset on the logarithmic scale, meaning there is a constant factor between them.

This is further illustrated by Fig. 10, which shows a comparison of measurements at 15° photon beam incidence made for the two positions on the sample with the four higher photon energies. The data shown in Fig. 10 corroborate that the measurement results for the two sample positions are offset by a constant factor. This is demonstrated by the green lines in Fig. 10. Here, the data represented by black symbols have been multiplied by a constant factor (as given in the figure caption) which bring them into agreement with the blue symbols (data measured at position 2). The same good agreement is found when the data measured at position 6 (yellow filled circles) are multiplied by a factor of three, giving the dark green dashed curve. This suggests that the shape of the emitted electron spectra is mainly determined by the nanoparticles themselves.

Fig. 11 shows a comparison of the spectra obtained on the 11 nm AuNP sample for measurement position 5 at 15° grazing incidence and for measurement position 3 at 60° incidence (in the plateau region of surface coverage [36], cf. Fig. S4.4) for all photon energies. The data measured at 12.0 keV for the two incidence angles appear to be offset by a constant factor (Fig. 11(b)), reflecting the different AuNP coverage at the respective measurement positions. For all other photon energies, large discrepancies in the relative

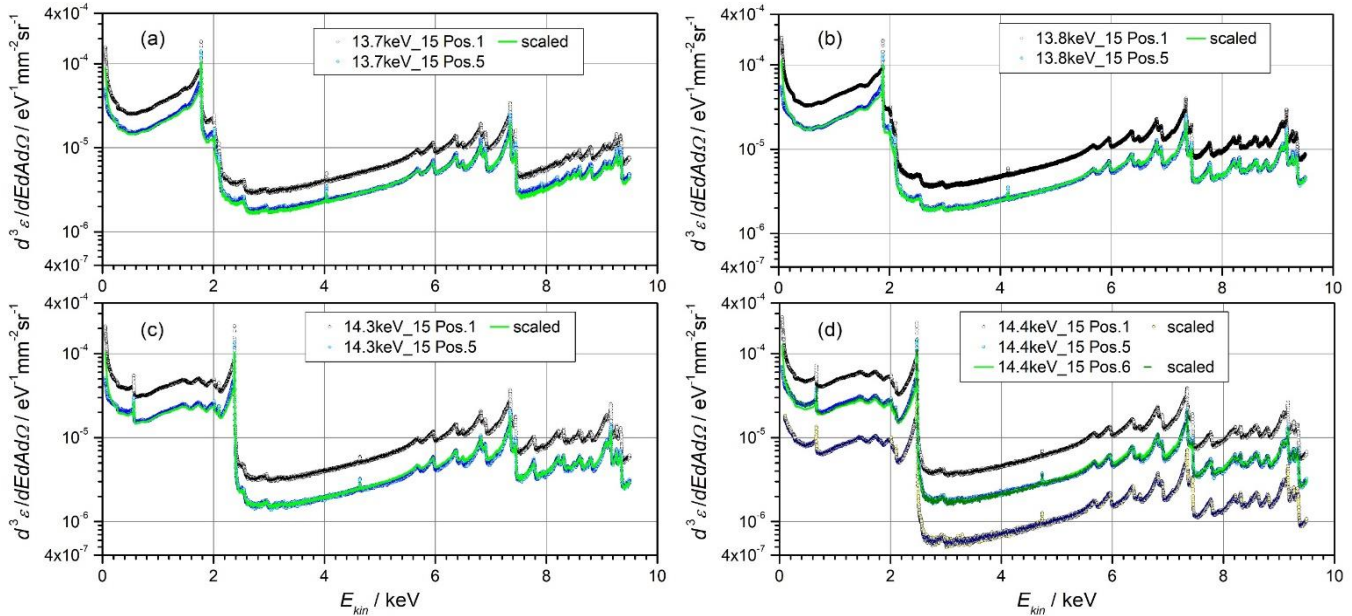

Fig. 10: Results for the spectral particle radiance per photon flux of the 11 nm AuNP sample measured at two (or three for 14.4 keV photons) positions on the sample with different AuNP coverage. The green lines show the modified values when the data indicated by the open black circles are multiplied by a constant factor of 0.573 (13.7 keV), 0.523 (13.8 keV), 0.486 (14.3 keV), or 0.466 (14.4 keV). The dashed dark green line indicates the modified values of the data marked by the filled yellow circles when multiplying by a factor of 3.

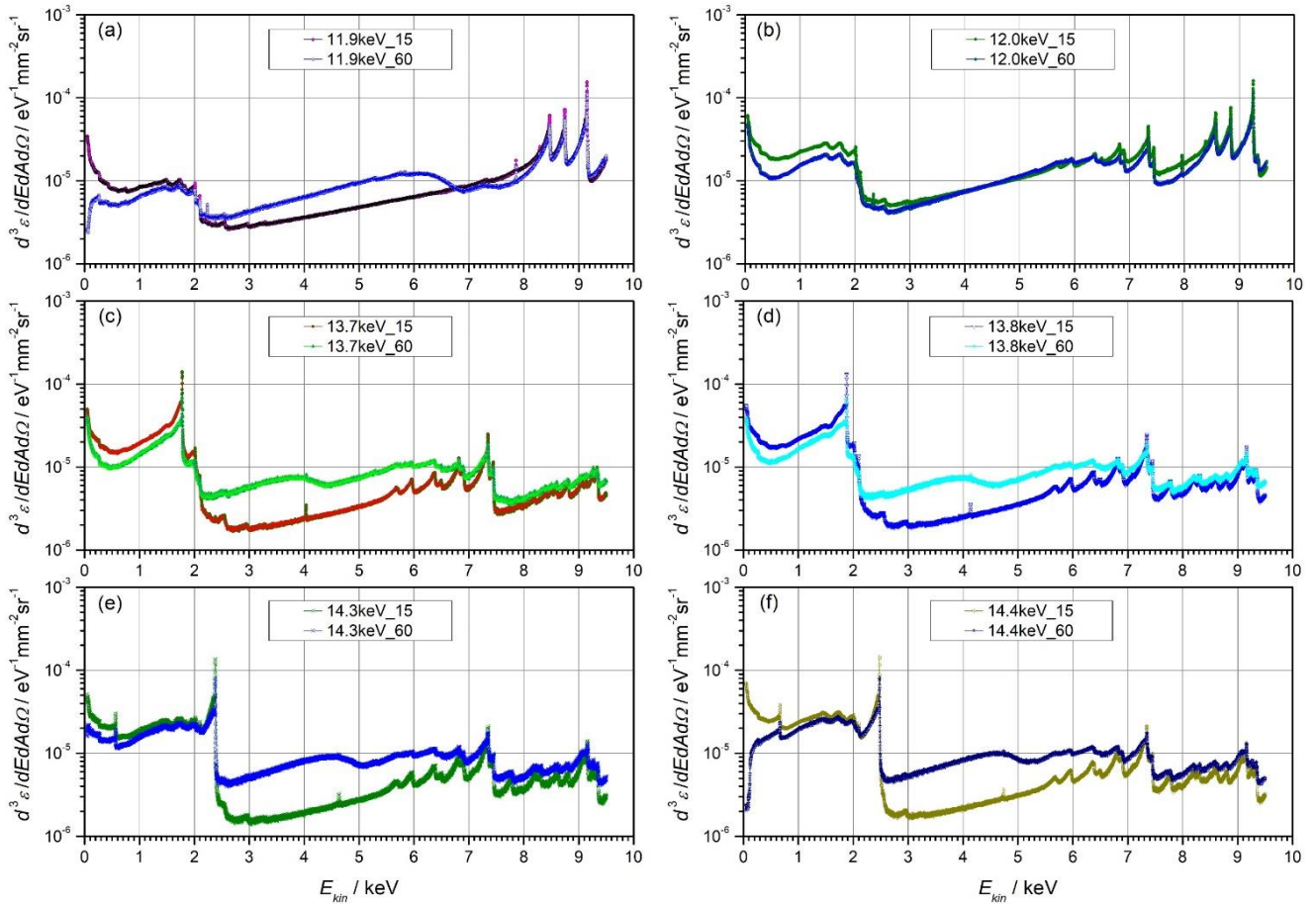

Fig. 11: Comparison of the data obtained on the 11 nm AuNP sample for 15° and 60° grazing incidence of the photon beam at measurement positions 5 and 3, respectively.

spectral-shape are evident in the electron energy range between 2.5 keV and 7 keV.

The origin of these discrepancies is a so-called Tougaard background [47,48] of electrons produced in the copper sample holder. This is supported by Fig. 12(a), which shows a comparison of the results on the 11 nm AuNP sample for 11.9 keV and 12.0 keV photons at 60° incidence with the single measurement on the bare carbon foil made at 14.4 keV and 15° incidence. The features appearing in the carbon foil data between 5.3 keV and 8 keV are also visible for the 11.9 keV data on the AuNP sample with similar relative shape. Their upper energies agree with the energy range of the Cu KLM and Cu KLL Auger electrons.

Fig. 12(b) shows the data for the other photon energies with the same carbon foil spectrum plotted against binding energy. It is evident that the feature observed between 9 keV and 11.5 keV binding energies are matched for the different photon energies, suggesting a photoelectron line as their origin. Their onset is at slightly higher energies than the Cu-K binding energy from the EADL.

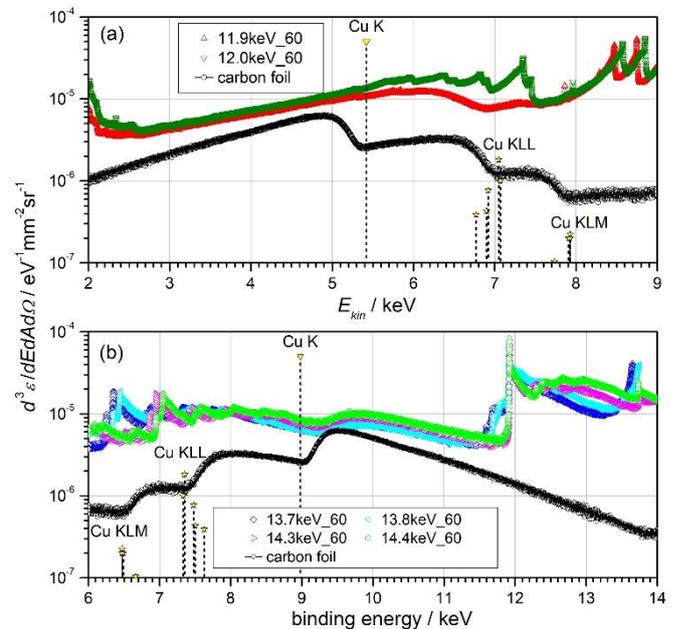

Fig. 12: Comparison of the results obtained with the 11 nm AuNP sample at 60° photon incidence and a measurement on the bare carbon foil plotted (a) versus kinetic energy and (b) versus binding energy. The energy positions of the Cu K photoelectron and the Cu KLL and KLM Auger lines refer to the measurement of the carbon foil at 14.4 keV photon energy.

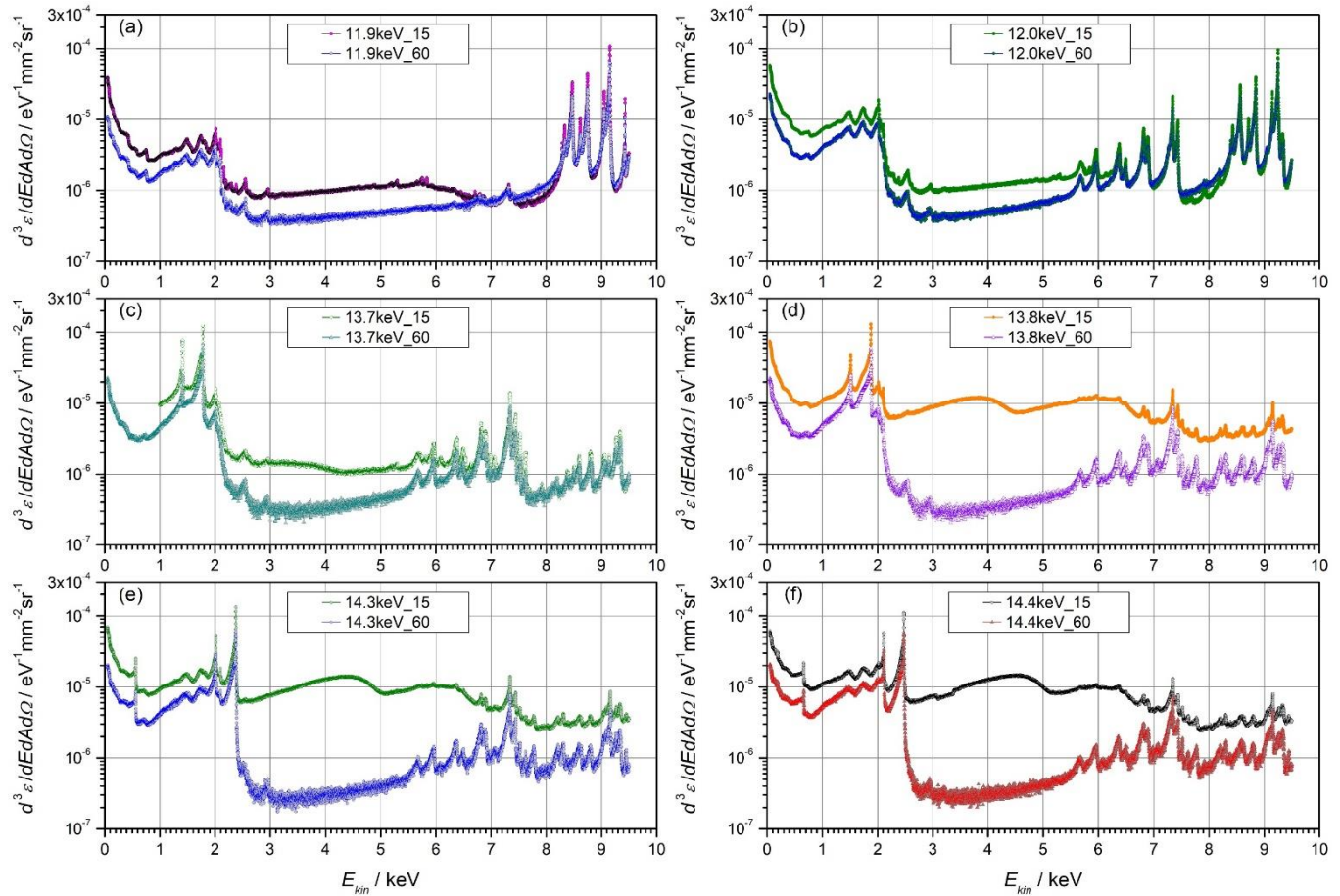

Fig. 13: Results for the spectral particle radiance per photon flux of the 5.3 nm AuNP sample measured at 15° and 60° incidence angles for the six photon energies. For the measurement at 13.7 keV and 15° incidence, the scans at lower energy were not recorded, so that this dataset contains only data from 1 keV kinetic energy.

### 3.5 Electron spectra of the 5.3 nm AuNP sample

For the sample with 5.3 nm AuNPs, only one measurement run was performed per incidence angle of the photon beam. The respective results are shown in Fig. 13, where large discrepancies between the measurements at the two incidence angles can also be seen. In contrast to the results obtained for the 11 nm AuNP samples, for this sample the Tougaard background is seen in the 15° experiments, while it does not appear to be present in the 60° measurements.

This finding is supported by the observation in Fig. S4.6 that the signal ratio of 15° to 60° at the chosen measurement position is about 6 for the energy of the Au L3 photoelectron line (Fig. S4.6(a) and (c)), while it amounts to more than 10 for the Au L3M5M5 Auger line (Fig. S4.6(b) and (d)).

## 4. Discussion

### 4.1 Tougaard background

The appearance of the Tougaard background in the experiments with the 11 nm AuNP sample is related to the fact that a 50 nm carbon foil covered with a (sub-)monolayer of AuNPs absorbs less than 0.6 % of the photons at a 15° angle of incidence and less than 0.2 % at 60°. At an incidence angle of 15°, the photon beam is expected to hit the sample holder on the aluminum support with a lateral offset (as seen from the spectrometer) of about 2.5 mm at a location shielded by the front edge (Fig. 14(a)). At an angle of incidence of 60°, on the other hand, the photon beam impinges on the copper block of the sample holder at a point about 1.2 mm from the spectrometer axis (Fig. 14(b)).

For the 5.3 nm AuNP sample, the Tougaard background is observed in the 15° measurements, but not at 60° incidence. One possible explanation for this unexpected observation is that in the 60° measurements on this sample, the photon beam hits a spot on the carbon foil that is on top of the aluminum support, so that the photon beam reaches the copper sample holder at a location below the aluminum layer (Fig. 15(b)).

In the 15° measurements, it is likely that the beam passed close to the edge of the aluminum support and hit the Cu sample holder below the self-supporting part of the carbon substrate (Fig. 15(a)). Since the carbon substrate has a curvature where it covers the aperture in the aluminum support (Fig. 4), the angle of incidence is reduced when the photon beam hits the sample near the circumference of the hole in the aluminum support. This implies that the irradiated surface area is enhanced by a factor which is greater than 1/sin(15°). This would explain why the 15° data are significantly larger than the 60° data for the three higher photon energies in the electron energy range above 8 keV, where electrons originating in the Cu sample holder cannot contribute to the measured signal. The magnitude of the discrepancies varying with photon energy can be attributed to slight shifts in beam position on the sample as the photon energy changes.

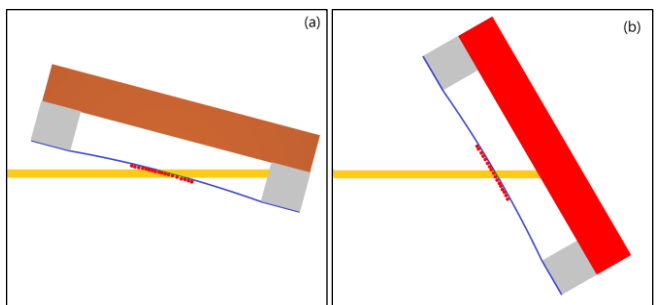

Fig. 14: Schematic representation of the irradiation scenario for the 11 nm AuNP sample at 15° (a) and 60° (b) photon-beam incidence. The orange thick line represents the photon beam, the red-brown rectangle represents the Cu sample holder, the gray rectangles represent the aluminum support, the blue line represents the carbon foil, and the red dots represent the AuNPs. The drawing is to scale with respect to the aspect ratio of the cavity behind the sample foil.

When the incident photon beam is dumped in the sample holder, the relaxation of core-ionized Cu or Al atoms can also produce fluorescence photons. Some of these photons interact with the AuNPs or the supporting carbon foil and generate electrons that contribute to the measured signal. The signal contribution originating from the Cu and Al fluorescence photons was estimated from the linear photon attenuation coefficients of the four materials (Au, carbon, Cu, Al) from the XCOM database [49]. The fluorescence photon energies and transition probabilities were obtained from the EADL database [39,40]. The photon interaction cross sections for the Au M-shells were taken from the database provided with the Penelope code [50].

Cu K$_\alpha$ photons occur with a probability of about 39 % and have energies slightly above 8 keV. Cu K$_\beta$ photons have energies slightly below 9 keV and occur with a probability of about 5 %. The energy of Al K$_\alpha$ photons is slightly below 1.5 keV, their production probability amounts to about 4 %.

For the case of irradiation shown in Fig. 14(a), the estimated ratio of the flux of Cu K$_\alpha$ and K$_\beta$ fluorescence photons to the flux of photons from the beamline was about 3.5×10$^{-4}$ and 4.5×10$^{-5}$. This estimate is based on the assumption that the fluorescence photons can produce detectable electrons over the whole area covered with AuNPs (about 1 mm²). The self-absorption of the fluorescence photons in the Cu material was taken into account, while the attenuation in Al was neglected.

This additional photon flux leads to the production of Au M-shell photoelectrons with energies between 4.6 keV and 5.8 keV for Cu K$_\alpha$ and between 5.6 keV and 6.8 keV for Cu K$_\beta$ photons. The estimated ratio between the peak intensity of these photoelectrons and of those produced by the incident photons is about 0.1 % for the Au M$_1$ line and between 0.15 % and 0.25 % for the Au M$_5$ line (at 11.9 keV and 14.4 keV photon energy, respectively). These values are negligibly small and explain why the corresponding peaks are not seen in Fig. 6.

In the geometry shown in Fig. 14(b), the solid angle subtended by the AuNP-covered surface is much larger (as seen from the region where the fluorescence photons are generated). In this case, the estimated flux ratio of Cu K$_\alpha$ and K$_\beta$ to the incident photons is about 1.3×10$^{-2}$ and 1.7×10$^{-3}$. The estimated ratio of Au M-shell photoelectrons produced by Cu fluorescence photons to those of the 11.9 keV beam ranges from 3.4 % for the Au M$_1$ line to 5.8 % for the Au M$_5$ line. At 14.4 keV photon energy, the corresponding ratio is 9.4 % for the Au M$_5$ line. Closer inspection of Fig. 11(a) shows that the Au M$_3$, M$_4$, and M$_5$ lines generated by Cu K$_\alpha$ photons can indeed be seen in this case.

A similar estimate was made for the signal from the carbon foil. Assuming that the relevant area was a factor of 10 larger than for the AuNP-covered surface, the ratio of photoelectrons produced in the carbon foil by fluorescence photons to those from the incident beam was about 2 % for an incidence angle of 15° and between 52 % and 82 % for the 60° geometry. However, the ratio of photoabsorption interactions in the carbon substrate to those in the AuNP sample is only about 2 %. Therefore, the increase in background signal from electrons generated in the carbon foil by Cu fluorescence is negligible.

For the additional contribution of Al K$_\alpha$ photons, an upper limit was estimated for the irradiation geometry shown in

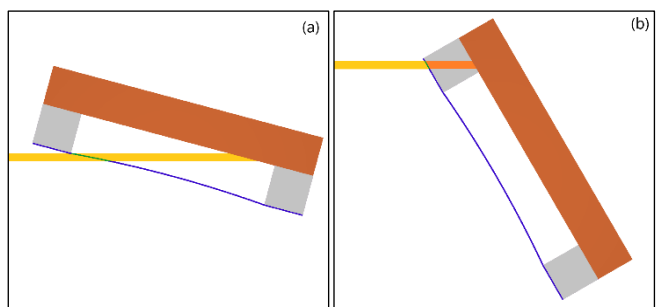

Fig. 15: Schematic representation of the presumed irradiation scenario for the 5.3 nm AuNP sample at 15° (a) and 60° (b) photon-beam incidence. The orange thick line represents the photon beam, the red-brown rectangle represents the Cu sample holder, the gray rectangles represent the aluminum support, and the blue line represents the carbon foil completely covered with AuNPs on the side facing the beam. The drawing is to scale with respect to the aspect ratio of the cavity behind the sample foil.

Table 3: Uncertainty budget of the spectral particle radiance per photon flux for the three samples studied.

| Source of uncertainty | Scope | 11 nm AuNP | | 5.3 nm AuNP | | Au foil |
| | | 15° | 60° | 15° | 60° | 15° |
|---|---|---|---|---|---|---|
| Electron spectrometer transmission | global | 33 % | 33 % | 33 % | 33 % | 33 % |
| Electron counting | data point | 2 % - 15 % | 3 % - 27 % | 3 % - 12 % | 7 %-30 % | 2 %-5 % |
| Energy range merging | spectrum | 0.8 % - 2.9 % | 1.1 % - 22 % | 0.7 % - 4.6 % | 1.5 % - 12 % | 0.9 % - 2.2 % |
| Energy scale | data point | 2 % - 10 % | 2 % - 10 % | 2 % - 10 % | 2 % - 10 % | 2 % - 10 % |
| Photon monitor measurement | data point | 0.6 % | 0.6 % | 0.6 % | 0.6 % | 0.6 % |
| Photon monitor calibration | spectrum | 2.5 % - 4.4 % | 2.5 % - 4.4 % | 2.5 % - 4.4 % | 2.5 % - 4.4 % | 2.5 % - 4.4 % |
| Sample inhomogeneity | sample | 5 % | 5 % | 10 % | 10 % | 0.5 % |
| Angle of incidence | sample | 5 % | 2 % | 5 % - 50 % | 2 % | 5 % |
| Photon beam area | global | 20 % | 20 % | 20 % | 20 % | 20 % |
| Total uncertainty | | 39 % - 43 % | 39 % - 53 % | 40 % - 66 % | 41 % - 53 % | 39 % - 41 % |

Fig. 14(a) by assuming that the photon beam is completely absorbed in aluminum. The resulting ratio of fluorescence photons to the incident beam was about $1.0 \times 10^{-5}$. The ratio of fluorescence photon-generated photoelectrons from the N, O, and P-shells of gold to incident photons was estimated to be 0.16%. The cross sections for photoabsorption on the outer shells of gold vary between 10 % and 5 % of the total photoabsorption cross section for photon energies between 11.9 keV and 14.4 keV. This means that the measured electron spectrum between 720 eV (energy of an Au $N_1$ photoelectron excited by Al $K\alpha$) and 1480 eV may contain a contribution of photoelectrons from outer Au shells that are produced by Al fluorescence. This is especially true in the case of irradiation geometry as shown in Fig. 15(b).

Proper modeling of the Tougaard background of electrons produced in the Cu and Al parts of the sample holder (and of the signal contributions from the Cu and Al fluorescence photons) is one of the challenges in the ongoing quantitative analysis of the spectra [51]. This will be presented in part 4 of the paper. The remainder of this discussion will focus instead on the plausibility of the results in terms of absolute magnitude, their associated measurement uncertainty, and their caveats.

### 4.2 Uncertainty budget

Table 3 shows an indicative uncertainty budget for the particle radiance measurement results considering only the dominant uncertainty contributions. The various sources of uncertainty are listed in the first column, and the scope of the respective uncertainty is stated in the second column. The term "scope" is used here in the sense that the uncertainty contribution applies to all data points within the respective scope, so that these contributions are fully correlated for these data points. Examples include the photon beam area and the spectrometer transmission for the reference energy and reference retarding ratio. This reference transmission value (cf. Supplement 3) is the same for all data points of all datasets and photon energies.

In contrast, the electron counting statistics are specific to each data point. As can be seen from the second row in Table 3, the respective uncertainties range between 2 % and 30 %. Larger uncertainties occur for the 60° measurements due to the smaller sample area (and, hence, the smaller number of AuNPs) irradiated by the photon beam.

The uncertainty component for electron range merging the spectra of different electron ranges is essentially the standard uncertainty of the ratio of transmission values for different retarding ratios determined from the measurements in overlapping energy intervals. The large range of uncertainties for the 60° measurements on the 11 nm sample are due to the two measurements at 14.4 keV, where there was a strong drift in the photon monitor signal. Omitting these two values reduces the upper limit of this uncertainty contribution to 4.2 %

The uncertainty contribution from variations of the energy scale of photons or electrons was estimated from the deviations from a constant value. Such deviations were seen in the ratios between different spectra measured for the same photon energy for the 11 nm AuNP sample (Fig. S4.3 and Fig. S4.5). These values were also used for the other samples.

The uncertainty of the photon monitor measurements was derived from a systematic review of the monitor current measurements in all experiments. The uncertainty of the photon monitor calibration depends on the photon energy (Table S1.1), and this uncertainty contribution is therefore correlated for all measurements at the same photon energy.

The uncertainty contribution due to sample non-uniformity was estimated from the position scans on the AuNP samples shown in Fig. S4.1, Fig. S4.4, and Fig. S4.6 for an estimated change in photon beam position of $\pm 50$ µm. For the gold foil, this uncertainty contribution was assumed to be an order of magnitude smaller than for the 11 nm AuNP sample, which still seems conservative.

For the 11 nm AuNP sample and the Au foil, the uncertainty contribution of the angle of incidence was estimated for the case where the true angle of incidence is within ±3° of the nominal value with a uniformly distributed probability. For the 5.3 nm AuNP sample, the values for the 60° measurements were estimated to be the same as for the other sample. And for the 15° measurements, the uncertainty was estimated to be up to 2.5 times higher (for the three higher photon energies), due to the assumed location of the irradiated area near the circumference of the hole in the aluminum support (Fig. 15(a)).

The row in Table 3 referring to the photon beam area only gives the uncertainty of the area determination (Supplement 2). The size of the photon beam is determined solely by the imaging optics and the size of the electron beam in the storage ring within the undulator and, therefore, the same for all photon energies.

As can be seen from the last row of Table 3, the total uncertainty is about 40 % for the 15° measurements on the 11 nm AuNP and the Au foil samples. Furthermore, the total uncertainty is up to more than 50 % for the 60° measurements on the two AuNP samples, and up to more than 60 % for some of the 15° measurements on the 5.3 nm AuNP samples. In the first case with the lowest total uncertainty, the two dominant uncertainty contributions come from the spectrometer transmission and the photon beam area. In the 60° measurements, the smaller signal gives rise to a large contribution from electron counting statistics which turns into the second most important uncertainty for some data points of these measurements.

### 4.3 Plausibility checks

The effective particle radiance is a quantity that characterizes the overall electron emission from the sample, including the emitted Auger and Coster-Kronig electrons as well as secondary electrons produced by inelastic electron interactions in the sample. It is the superposition of the particle radiance from the actual sample, $d^3\varepsilon_s/dAd Ed\Omega$, and the background contribution, $d^3\varepsilon_b/dAdEd\Omega$.

$$\frac{d^3\varepsilon}{dA\,dE\,d\Omega} = \frac{d^3\varepsilon_s}{dA\,dE\,d\Omega} + \frac{d^3\varepsilon_b}{dA\,dE\,d\Omega} \tag{3}$$

For the AuNP samples, a more relevant quantity is the yield of emitted electrons per photon interaction in a nanoparticle, $d^2y_{np}/dEd\Omega$. This quantity is doubly differential in energy and solid angle and is given by

$$\frac{d^2y_{np}}{dE\,d\Omega}(E_e) = \frac{1}{n_i(E_p)}\frac{d^3\varepsilon_s}{dA\,dE\,d\Omega} \tag{4}$$

where $n_i$ is the mean area number density of AuNPs undergoing a photon interaction. $n_i$ is given by

$$n_i(E_p) = p_i(E_p) \times c_s \tag{5}$$

where $p_i$ is the probability of an incident photon of energy $E_p$ to interact with an AuNP, and $c_s$ is the number of AuNPs per surface area (surface number density).

The probability $p_i$ can be estimated from the known particle size, beam size, and literature data of the photon interaction cross sections as follows

$$p_i(E_p) = \frac{\mu_{tot}(E_p)V_{np}}{A_b} \tag{6}$$

where $\mu_{tot}$ is the total linear attenuation coefficient, $V_{np}$ is the volume of the AuNP, and $A_b$ is the beam cross-sectional area.

In the evaluation of Eq. (6), the average volume was used. This was calculated assuming a Gaussian distribution of the AuNP diameters. A standard deviation of 0.9 nm was used for the 11 nm AuNPs. (The value of 1.8 nm given in [36] corresponded to two standard deviations.) For the 5.3 nm AuNPs, it was assumed that the relative standard deviation of the diameter distribution was the same as for the two types of smaller AuNPs reported in [52]. This resulted in a standard deviation of 1.5 nm for the AuNP diameter.

The linear attenuation coefficients were obtained by adding the requested energies in the online search form of the XCOM database [49]. The resulting values of $p_i$ for the two nanoparticle types are listed in the second and fourth columns of Table 4.

The values of the product $n_i = p_i \times c_s$ are shown in the third and the last columns of Table 4 for the surface density of a hexagonal close-packed (HCP) monolayer of mean-diameter AuNPs with a 2 nm thick PEG or PEG-MUA coating. The respective values are $c_s = 5.13 \times 10^9$ mm$^{-2}$ for the 11 nm AuNPs and $c_s = 1.34 \times 10^{10}$ mm$^{-2}$ for the 5.3 nm AuNPs.

When the doubly differential yield of emitted electrons per photon interaction in a nanoparticle was known for all emission angles and the complete energy range, the total number of electrons emitted per photon interaction in the AuNP could be calculated by integrating over energy and the whole solid angle. Calculating the integral after multiplying the yield by the electron energy would give the average energy per photon interaction that is transported out of the nanoparticle by electrons generated in it. These electrons are produced by photoelectric absorption, by atomic relaxation after photoabsorption, and by interactions of the electrons emitted from the absorbing atom in the AuNP.

The calculation of these quantities from the experimental data is hampered by the fact that the background particle radiance $d^3\varepsilon_b/dA\,dE\,d\Omega$ and the angular distribution of the emitted electrons are not known *a priori*. Moreover, due to the limited energy range of the HAXPES spectrometer, only a part

Table 4: Probability per incident photon of the two types of AuNPs to experience a photon interaction and its product with the average surface area density $c_s$ of AuNPs for a hexagonal close-packed monolayer.

| Photon energy (keV) | 5.3 nm AuNP | | 11 nm AuNP | |
|---|---|---|---|---|
| | $p_i$ | $n_i$ (mm$^{-2}$) | $p_i$ | $n_i$ (mm$^{-2}$) |
| 11.9 | $1.17\times10^{-12}$ | 0.016 | $8.63\times10^{-12}$ | 0.048 |
| 12.0 | $2.81\times10^{-12}$ | 0.038 | $2.08\times10^{-11}$ | 0.116 |
| 13.8 | $1.97\times10^{-12}$ | 0.026 | $1.46\times10^{-11}$ | 0.081 |
| 13.9 | $2.67\times10^{-12}$ | 0.036 | $1.98\times10^{-11}$ | 0.110 |
| 14.3 | $2.45\times10^{-12}$ | 0.033 | $1.82\times10^{-11}$ | 0.101 |
| 14.4 | $2.78\times10^{-12}$ | 0.037 | $2.06\times10^{-11}$ | 0.115 |

of the electron energy spectrum could be detected in the experiments.

Nevertheless, estimates can be made by replacing $d^3\varepsilon_s/dAdEd\Omega$ in Eq. (4) with $d^3\varepsilon/dA\,dE\,d\Omega$ and using the values of $n_i$ for a monolayer coverage (Table 4). Integrating this estimate for $d^2y_{np}/dE\,d\Omega$ over the measured energy range and multiplying it by $4\pi$ (i.e., assuming isotropic emission) gives the values listed in Table 5.

Several observations can be made about Table 5. One is that the values obtained in measurement positions 1, 2, and 4 are approximately the same for all photon energies. That positions 2 and 4 give the same value as position 1 is surprising, since they were located regions of lower surface coverage (Fig. S4.1(d)). However, this paradox could be due to a coincidence, since for positions 3 and 5, the values at the lower photon energies are significantly increased compared to the higher photon energies.

The values obtained at position 5 for the four higher photon energies are about half those for position 1, which is expected from what is shown in Fig. 10. Similarly, the factor of 3 between the values for 14.4 keV at positions 5 and 6 are also expected from the results shown in Fig. 10(d). It is worth noting that for both position 3 and position 5, there is a decreasing trend for the three lower photon energies and essentially constant values for the three higher photon energies.

The fact that for 14.4 keV, the value at position 3 is about 20 % higher than the value at position 5 – even though both positions are in the plateau region – is explained by the extra electrons from the Tougaard background present in the measurements at position 3.

For the 5.3 nm AuNP sample, the values for 60° are more or less independent of photon energy, whereas clear differences between photon energies are found for the 15° measurements.

The data shown in Table 5 are the estimated number of electrons with energy less than 9.5 keV emitted per AuNP undergoing photon interaction. They mainly reflect the differences between measurement positions and incidence angles already seen in the figures. Therefore, a more relevant key figure is obtained by integrating the product of the estimated $d^2y_{np}/dE\,d\Omega$ and the electron energy over the measured energy range, multiplying the result by $4\pi$ and dividing by the photon energy. This quantity is an estimate of the proportion of the energy transferred by the photon interaction that is transported out of the AuNP by electrons with kinetic energies below 9.5 kV. This quantity is listed in Table 6.

While the same observations as in Table 5 can be made about trends with photon energy and ratios between different columns, the values now allow quantitative interpretation. For an isolated AuNP, this quantity must be significantly smaller than unity. This is because the integral is performed over only a part of the electron energy spectrum and the total energy transported out of the AuNP cannot exceed the photon energy. Ad hoc simulations with the Penelope code [50,53] forcing photon interaction in the AuNPs indicated that about 70 % of the photon energy was found in the energy spectrum of escaping electrons with energies below 9.5 keV.

For AuNPs in a monolayer on a thin carbon substrate, the number of electrons emitted from the sample in the direction of the electron spectrometer is enhanced by the scattering of emitted electrons on neighboring AuNPs. For an HCP monolayer of 5.3 nm AuNPs with a 2 nm PEG coating, the solid angle covered by the first and second nearest neighbors is about 2 sr or 16 % of 4π. In addition, for the 11 nm AuNPs, the respective value is 27 % of the full solid angle. However, these nanoparticles contain eight times more gold atoms than the smaller AuNPs, resulting in correspondingly stronger electron scattering. Moreover, STEM investigations of the 11 nm AuNP sample showed a significant occurrence of multilayers [36], where the coordination number of the nearest neighbors can be a factor of 2 higher than for an HCP monolayer.

Table 5: Integral of the electron yield per photon interaction over the measured electron energy range multiplied by $4\pi$ sr for the particle radiance obtained for the different photon energies at the different measurement positions on the two AuNP samples. The values have been calculated assuming a hexagonal close-packed monolayer of AuNPs.

| Photon energy (keV) | Measurement position on 11 nm AuNP sample | | | | | | 5.3 nm AuNP sample | |
|---|---|---|---|---|---|---|---|---|
| | 1 | 2 | 3 (60°) | 4 | 5 | 6 | 15° | 60° |
| 11.9 | - | - | 24.6 | 18.4 | 23.1 | - | 6.4 | 4.0 |
| 12.0 | - | 19.2 | 15.1 | - | 17.4 | - | 4.3 | 2.7 |
| 13.7 | 20.1 | - | 14.2 | - | 12.1 | - | 4.6 | 3.4 |
| 13.8 | 19.4 | - | 11.5 | - | 10.2 | - | 10.7 | 2.7 |
| 14.3 | 20.5 | - | 12.1 | - | 10.1 | - | 11.3 | 2.8 |
| 14.4 | 22.0 | - | 11.4 | - | 10.6 | 3.5 | 10.2 | 2.9 |

Table 6: Ratio of the integral of the electron yield per photon interaction multiplied by the electron energy to the photon energy for the particle radiance obtained for the different photon energies at the different measurement positions on the two AuNP samples.

| Photon energy (keV) | Measurement position on 11 nm AuNP sample | | | | | | 5.3 nm AuNP sample | |
|---|---|---|---|---|---|---|---|---|
| | 1 | 2 | 3 | 4 | 5 | 6 | 15° | 60° |
| 11.9 | - | - | 12.39 | 8.82 | 11.56 | - | 2.36 | 1.76 |
| 12.0 | - | 7.97 | 6.77 | - | 7.43 | - | 1.31 | 1.02 |
| 13.7 | 4.79 | - | 4.20 | - | 2.89 | - | 1.01 | 0.65 |
| 13.8 | 4.73 | - | 3.43 | - | 2.50 | - | 2.84 | 0.54 |
| 14.3 | 4.55 | - | 3.38 | - | 2.21 | - | 2.87 | 0.54 |
| 14.4 | 4.66 | - | 3.14 | - | 2.20 | 0.73 | 2.53 | 0.54 |

Thus, ignoring the data affected by the Tougaard background, the numbers in Table 6 indicate that for the 11 nm AuNPs, electron scattering on neighboring AuNPs results in a significant enhancement of the number of electrons measured with the spectrometer. Moreover, the relative magnitude of this enhancement depends on the surface coverage and the presence of multilayers.

For the measurements at position 3 and 5 of the 11 nm AuNP sample and for 60° incidence on the 5.3 nm AuNP sample, the trend of increasing values with decreasing photon energy is more pronounced in Table 6 than in Table 5. The weighting by electron energy applied in deriving the data in Table 6 implies that these values are predominantly due to the M-shell photoelectron peaks in the measurements at 11.9 keV and 12.0 keV photon energy.

In contrast to Auger electrons, the emission direction of photoelectrons has a non-isotropic distribution peaking along the direction of the electric field vector of a linearly polarized photon beam. The respective anisotropy can be described in terms of Legendre polynomials [54,55]. For excitation from an s-orbital of an isolated atom, the number of electrons ejected along the direction of the electric field vector is higher than for isotropic emission by a factor of 3. However, the $M_3$, $M_4$, and $M_5$ lines originate on p-orbitals for which this enhancement is smaller.

Therefore, the anisotropy of photoelectron emission cannot fully explain the pronounced increase of the values listed in Table 5 and Table 6 with decreasing photon energy. It should also be noted that the peculiarities in the variation of the photon monitor calibration factor and photon flux

(Table S1.1) do not provide an explanation either. Due to the energy dependence of the photon interaction coefficients, higher values of the calibration factor are expected at lower photon energies. However, higher calibration factors would lead to even larger values of the figures of merit at the two lowest photon energies.

The larger contribution of M-shell photoelectrons in the spectra at the lowest photon energies could also provide another possible explanation. These electrons have energies in the range between 5 keV and 9.5 keV, giving them a range of a few hundred nm in gold. Neighbors within this range could therefore contribute to the measured signal by electron scattering. What argues against this explanation is that the ratios of the values at position 1 (and position 6) to those at position 5 of the 11 nm AuNP sample are approximately the same in both Table 5 and Table 6.

The most likely reasons for the increased values at the lower two photon energies are as follows: First, that the position of the photon beam on the sample varies with photon energy, with the beam with lower photon energies hitting a spot with higher AuNP coverage. Second, that there is probably a Tougaard background from Al K photoelectrons. It can be seen from Fig. 12(a) that this background contributes most strongly for Cu in the energy range between 100 eV and 2 keV below the K-shell photoelectron energy. Assuming a similar relationship for the Al photoelectrons and considering that the K-shell binding energy of Al is about 1.56 keV, this background would affect the signal in the energy range of above 8 keV. This applies to the electron spectra measured at 11.9 keV and 12.0 keV photon energy, but not at higher photon energies.

In terms of absolute numbers, the values in the last column of Table 6 suggest that the 5.3 nm AuNPs do not form an HCP monolayer and/or have a coating thickness larger than the assumed 2 nm. For the 11 nm AuNP sample, position 6 would be a region of HCP monolayer coverage, and position 5 would already have multilayers. This is in contradiction to the estimated average AuNP coverage in the order of 0.2 monolayers reported in [36]. It must be noted, however, that this estimate of coverage was based on the nominal concentration of the AuNP solution and is not a result from the STEM measurements. Furthermore, the area covered by the AuNPs that was estimated based on the linear dimensions read from Fig. S4.4 is only about 2/3 of the value determined from the STEM images in [36], which suggests that the average coverage was higher by about 50 % for the sample used in this study.

An approximate quantitative analysis of the STEM images suggests that some of the AuNPs have the shape of prolate ellipsoids standing with the long axis (elevated by about 20 % relative to the short axis) pointing upward. In addition, the preliminary characterization of the AuNP solution by small angle X-ray scattering (SAXS) indicates a significant fraction of about 30 % of AuNP dimers and trimers and an increased average volume of the AuNPs by about 25 % compared to the STEM values. All these factors suggest the actual probability for photon interaction could be increased by a factor of about 1.6. Even after correcting the values for the 11 nm AuNP sample in Table 6 with this increased photon interaction probability, the conclusion still remains, that only the measurement at position 6 was for (sub-)monolayer coverage. All other measurement spots were, however, in regions with significant multilayer coverage.

### 4.4 Influence of surface contamination

The experiments were conducted on the samples without any further surface cleaning procedures after transferring them to the vacuum system of the experimental station. The reason for this was that previous studies had shown that treating the samples with ion beam sputtering can lead to the destruction of the AuNP coating or the removal of the AuNPs from the sample surface [37]. The setup did not allow the use of the discharge etching method described in [46] either. Therefore, it must be assumed that the samples are covered with surface impurities due to the adsorption of residual gas atoms as well as residues of the chemicals used in their preparation, as already seen in Fig. 6. The XPS investigations on similar 11 nm AuNP samples showed surface coverage with hydrocarbons and other carbon-based molecules [37].

As Henneken et al. have shown, surface contamination can drastically reduce the electron emission properties of metal surfaces under soft X-ray irradiation [42–44]. Hespeels et al. also found a strong suppression of low-energy electron emission from AuNP samples due to carbon surface contamination [46]. Evidence for this can also be seen in the 60° data in Fig. 11(a).

Therefore, the measured electron radiance at low kinetic energies may be underestimated due to surface contamination. This could be another reason why the two sets of measured data at position 3 of the 11 nm AuNP sample show poor reproducibility in the energy range up to a few hundred eV (cf. Fig. S4.3 and Fig. S4.5). In fact, Fig. S4.3(c) to (f) suggest that surface contamination is more prevalent in the regions with monolayer coverage. This is not confirmed by Fig. S4.5, which shows a higher average signal associated with a reduction at low energies. This indicates that not only the surface density of AuNPs, but also the density of the surface contamination is strongly non-uniform. However, it is important to note that this contamination significantly affects the measured particle radiance only at lower electron energies, as was also found in [46]. Conversely, this means that the results at higher energies can very well serve as benchmark data for simulations.

### 4.5 Use of the results for benchmarking

In the first report by Li et al. [18] on the results of the code comparison, discrepancies between electron spectra reported by different participants were in some cases above one order of magnitude. The literature review by Moradi et al. [16] also included examples of order-of-magnitude differences between the results of various MC simulation studies of nanoparticle radiation effects.

Given this situation, the fact that the experimental dataset produced in this study is only for six photon energies and is subject to large uncertainties does not seem to impair its use as a benchmark. This is simply because the spread of results in the literature is much larger. Therefore, tables of the results are provided as Supplements 5 to 10 to this article.

During the re-evaluation of the results from the multicenter comparison, which led to the publication of a corrigendum [19], it was discovered that two participants had used the cumulative distribution of the X-ray spectra as the photon spectrum instead of the probability distribution. At the same time, these participants also used a different photon beam size. The synergy of effects resulting from the two deviations from the exercise definition made it very difficult to identify the origin of the deviations of the results of these two participants from the others. Reference simulations for a fixed photon energy would have made it easier to identify the geometry problem.

The measured data refer to two specific nanoparticle sizes with particular coatings and layer thicknesses. Therefore, when using the data for benchmarking simulations in studies interested in different sizes of nanoparticles and/or coating thicknesses, two simulations are required: one for the sizes used here and one for the sizes to be studied. Once set up, changing dimension and/or material compositions in a simulation geometry is not too much of an effort and should be less error-prone.

Within certain limits, the measured data can also be used directly for other nanoparticle sizes and coating thicknesses. The results shown in Fig. 9 and Fig. 10 indicate that spectra obtained without significant background contributions are very similar. Moreover, there is also a qualitative similarity of these data and the data shown in Fig. 13 for the 60° measurements on the 5.3 nm AuNP sample. Therefore, it is expected that the electron spectra of larger or smaller nanoparticles (after interaction with photons of the energies considered here) can be approximated by scaling with the interaction probability of the photons in the nanoparticles according to Eq. (1).

The approximation outlined in the previous paragraph should be applicable in the electron energy range above about 3 keV, that is, for Au L Auger electrons and the partially decelerated photoelectrons of the M and higher shells of gold. Within the uncertainties of the present experiment, it may also be reasonably applicable at lower electron energies, although caution should be taken when considering different coating thicknesses or materials (or even a coating plus conjugated biomolecules). A study by Morozov et al. [56] showed a 51 % suppression of low-energy electrons in the energy range below 3 keV for a PEG coating thickness of 8.5 nm. This thickness is comparatively large compared to the values reported in the review by Kuncic and Lacombe [6]. Taking the value reported by Morozov et al. as a reference, one would roughly expect an attenuation of low-energy electrons by about 12 % due to the presence of a PEG coating of 2 nm.

All of the preceding arguments apply in the case of the photon energies used in this study. The experiments reported here for only six photon energies were performed during a four-day beamtime shift. Therefore, measuring electron emission data for a wider range of photon energies is not possible given the limited availability of beamtimes at high-intensity synchrotron radiation facilities. However, photon interaction cross sections are known with uncertainties in the range of a few percent [57]. In principle, it should thus be possible to construct benchmark datasets derived from the present measurements using the XCOM photon interaction database [49]. This requires a decomposition of the measured spectra into the components that change with photon energy (photoelectrons) and those whose energy is independent of photon energy (Auger electrons). This decomposition will be the subject of the fourth part of the paper, where a detailed line shape analysis of the measured spectra will be presented.

## 5. Conclusions

In this work, experiments were performed on AuNP samples with a data analysis aimed at obtaining absolute results for the particle radiance of emitted electrons over a wide range of kinetic energies after photoabsorption on the L-shells of gold. To the best of our knowledge, comparable data were available in the literature only for lower energy photons after excitation of gold M-shells [58] and for irradiations with protons [52], in both cases only on a relative scale. Experimental benchmarks of numerical methods for determining dose enhancement by high-Z materials have been performed by Alawi et al. [22], Mirza et al. [23] and Gray et al. [24]. However, the respective experiments do not allow conclusions to be drawn about the energy spectra of emitted electrons as they determined the integral dosimetric effect.

The absolute determination of electron radiance proved very challenging because the major uncertainty components arise from the transmission of the electron spectrometer and the cross-sectional area of the photon beam. A photon flux monitor was used to correct for temporal changes in the photon flux and proved essential for identifying transients following changes of the photon energy. To obtain absolute values of photon flux, the photon flux monitor was calibrated. A strong variation of the calibration factor and photon flux with photon energy was found, which originated in the use of the wrong coating on the last focusing mirror. From the plausibility checks presented here and the results on the gold foil sample that will be presented in the second part of the paper, it can be ruled out that this peculiarity of the experiment impairs the results.

A major drawback of the experimental results is the non-uniform surface coverage with AuNPs and the Tougaard background of electrons produced in the sample holder. However, the fact that the emitted electron spectra were studied over large energy ranges using different samples and irradiation geometries allowed a consistent understanding of the limitations of the results.

Potential future repetitions of these experiments with the aim of producing data with lower uncertainties will need to reduce the two uncertainty contributions of global scope. For the photon beam area, this could be achieved by using a specially designed test sample that has spatial structures of about 10 μm dimensions and the possibility to study the beam in a wide range of azimuthal angles. For the electron spectrometer transmission, a reduction of uncertainty can only be reasonably achieved by electron transport simulations that take into account the details of the measurement geometry and the emitted electron spectra.

To obtain better quality data, more uniform samples and the efficient suppression of the Tougaard background will be essential. The samples used in the present experiments were designed to serve multiple purposes, including the possibility to perform TEM and low-energy proton measurements. Diamond plates with thicknesses in the micrometer range may be considered for future hard X-ray photoelectron spectroscopy measurements. These would be opaque to the electrons of a copper sample holder, and the photoelectron background should be smooth due to the low binding energy of electrons in carbon. The mechanical stability of such diamond plates would allow the wet chemical deposition of AuNPs in a more uniform monolayer [36]. AuNP samples with more uniform and known surface density will allow the determination of the radiance per AuNP rather than per area, eliminating the need to know the cross-sectional area of the beam and thus removing an important source of uncertainty.

In the present experiments, the instrumentation at the experimental station was used for the first time to measure electron emission spectra over the entire available energy range at higher photon excitation energies than commonly used. Optimizing the experimental sequences to mitigate the effects of transients in the photon beam on the measurements could then further reduce the experimental uncertainties, so that overall uncertainties of the determined particle radiance to values on the order of 10 % seem achievable.

For the foil sample, some of the caveats identified for the AuNP samples are less relevant, as will be shown and further discussed in the second part of the paper. Despite these limitations, the results on the nanoparticle samples are still suited for further analysis such as benchmarking theoretical investigations and detailed line fitting analysis as in conventional XPS studies. These analyses are presented in the third and fourth parts of the paper.

## 6. Acknowledgements

This work was funded by the German Research Foundation (DFG) under grant number 386872118. DESY (Hamburg, Germany), a member of the Helmholtz Association HGF, is acknowledged for the provision of experimental facilities. Parts of this research were carried out at PETRA III, and the authors would like to thank Patrick Lömker for assistance in using beamline P22 with the HAXPES setup. Beamtime was allocated for proposal I-20200068. The authors would also like to thank Benedikt Rudek for initiating this project, Elisabetta Gargioni, Florian Schulz, Stéphane Lucas, and Félicien Hespeels for providing the gold nanoparticle samples, Dietmar Eberbeck for performing the SAXS measurements, Heike Nittmann for producing the drawing of the experimental setup, Ella Jones for linguistic proofreading of the manuscript, and Andreas Pausewang for manufacturing the aluminum supports of the samples and for support during the beamtime. SPECS Surface Nano Analysis GmbH is gratefully acknowledged for performing the simulations of the electron spectrometer transmission.

## 7. Author contributions:

**HR:** Methodology, Validation, Data Curation, Formal analysis, Supervision, Visualization, Writing - Original Draft, Writing - Review & Editing **PAH:** Sample Preparation, Investigation, Writing - Review & Editing **AH, CS:** Investigation, Writing - Review & Editing **SL:** Providing Samples, Writing - Review & Editing **WYB:** Investigation, Supervision, Writing - Review & Editing

## Supplement 1: Photon flux calibration

To calibrate the photon flux monitor, additional measurements were taken with a calibrated AXUV100GX photodiode at the sample position. The angle of incidence was 45° and the current of the photodiode was measured with a different channel of the same CAEN instrument used to measure the current from the photon flux monitor. Dark current suppression was used, resulting in residual dark currents below 10 pA for both, the photodiode and the photon flux monitor. The measured photocurrent of the diode ranged from 50 µA to 750 µA, depending on the photon energy and the attenuators used.

Prior to the measurements, the photon beam was scanned alternately horizontally and vertically across the photodiode to determine the center position. Measurements were then taken at two sets of 20 equidistant positions that varied between − 1 mm and + 1 mm from the center position in either the horizontal or vertical direction. The dwell time per data point and the integration time for both channels was 1.0 s.

The calibration factor $Q_0(E_p)$ of the photon flux monitor at photon energy $E_p$ was then determined using Eq. (S1.1):

$$Q_0(E_p) = \langle I_0/I_d \rangle \times s_d(E_p) \times E_p \qquad (S1.1)$$

Here $\langle I_0/I_d \rangle$ is the average over all measurement positions of the ratio of the measured currents $I_0$ of the photon flux monitor and $I_d$ of the photodiode. The spectral responsivity, $s_d$, of the photodiode was obtained by extrapolating the data from the calibration certificate using a dead-layer model according to Eq. (S1.2).

$$s_{d,cal}(E_p) = \frac{\mu_{en}}{W \times \mu}(1 - e^{-\mu d_1})e^{-\mu d_2} \qquad (S1.2)$$

In Eq. (S1.2), $W = (3.64 \pm 0.03)$ eV is the mean energy required to produce an electron-hole pair in silicon [1], and $d_1$ and $d_2$ are the two model parameters, namely the thickness of the active layer and the dead layer, respectively. (The model assumes complete charge collection in the active layer and no charge collection in the dead layer).

The quantities $\mu_{en}$ and $\mu$ are the linear energy-absorption and linear attenuation coefficients of silicon, respectively. Both coefficients depend on the photon energy and were calculated from the mass energy absorption coefficient $\mu_{en}/\rho$ and the mass attenuation coefficient $\mu/\rho$ using the mass density of silicon $\rho = (2.3296 \pm 0.0001)$ g/cm³. The values for $\mu/\rho$ were obtained from the NIST XCOM database [2] by performing a database query for the desired energies. Data for $\mu_{en}/\rho$ were obtained from the NIST XMAADI database [3], and a generalized spline interpolation [4] was used to obtain the values at the desired energies.

The model parameters and $d_1$ and $d_2$ were determined by fitting (S1.2) to the data given in the calibration certificate (for photon energy values between 3 keV and 10 keV) using the generalized reduced gradient method implemented in the Excel solver. The resulting parameter values for $d_1$ and $d_2$ were $(52.7 \pm 0.3)$ µm and $(0.04 \pm 0.05)$ µm, respectively. The fit of the model curve to the calibration data points is shown in Fig. S1.1.

Since the photodiode calibration was performed for normal incidence, while the photons were incident at 45° during the measurements, the spectral responsivity was calculated using Eq. (S1.3).

$$s_d(E_p) = \frac{\mu_{en}}{W \times \mu}(1 - e^{-\mu\sqrt{2}d_1})e^{-\mu\sqrt{2}d_2} \qquad (S1.3)$$

Table S1.1: Results of the calibration of the photon flux monitor. The uncertainties are given for a coverage factor $k = 1$. The relative uncertainty of the photon energy is about $1.25 \times 10^{-4}$ ($k = 1$). The average measured currents of the monitor and the standard deviations are shown in the third column for the case when no attenuators were used. Therefore, the photon flux given in the last column is the maxium value that was generally used for the measurements with the nanoparticle samples.

| Photon energy / keV | $Q_0$ / nAs | $I_0$ / nA | $N_p$ / s$^{-1}$ |
|---|---|---|---|
| 11.9 | $(0.80 \pm 0.02) \times 10^{-11}$ | $37.9 \pm 0.8$ | $(4.77 \pm 0.16) \times 10^{12}$ |
| 12.0 | $(0.89 \pm 0.04) \times 10^{-11}$ | $39.5 \pm 1.1$ | $(4.42 \pm 0.23) \times 10^{12}$ |
| 13.7 | $(2.73 \pm 0.08) \times 10^{-11}$ | $37.9 \pm 0.9$ | $(1.39 \pm 0.05) \times 10^{12}$ |
| 13.8 | $(2.58 \pm 0.10) \times 10^{-11}$ | $36.5 \pm 0.9$ | $(1.41 \pm 0.06) \times 10^{12}$ |
| 14.3 | $(2.58 \pm 0.08) \times 10^{-11}$ | $30.6 \pm 0.5$ | $(1.18 \pm 0.04) \times 10^{12}$ |
| 14.4 | $(2.28 \pm 0.06) \times 10^{-11}$ | $28.3 \pm 1.1$ | $(1.24 \pm 0.06) \times 10^{12}$ |

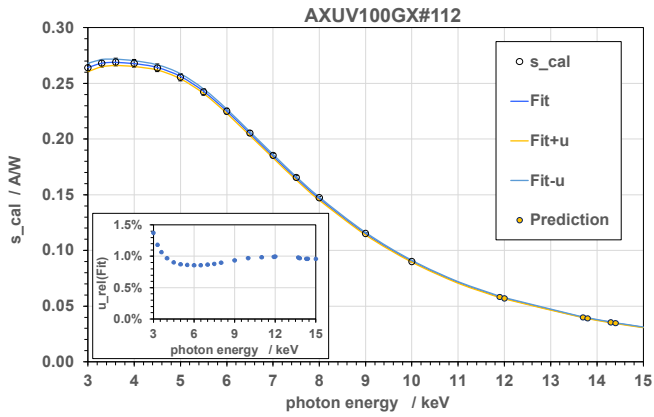

Fig. S1.1: Photodiode calibration curve.

The responsivity values calculated using Eq. (S1.3) were independently confirmed by simulations of the absorbed energy in the sensitive layer of the photodiode. These simulations were performed with Penelope (release 2006) using a custom-built main program that also allowed the influence of an electron emission pattern based on the linear polarization of the radiation to be evaluated. The difference between a $\cos^2\theta$ distribution (where $\theta$ is the polar angle with respect to the polarization vector) and an isotropic distribution was on the order of 0.1 %. The spectral responsivity calculated from the average absorbed energy per photon deviated from the values obtained by Eq. (S1.1) by about 1.2 % which is within the uncertainties of the interaction cross sections [5]. The deviations were added as additional uncertainty contributions to the uncertainties of the fit curve obtained by uncertainty propagation, which are shown in the inset of Fig. S1.1.

The photon flux monitor calibration factors are listed in the second column of Table S1.1. The third column shows the mean and the standard deviation of the recorded photocurrents from all measurements without attenuators in the beam, and the last column shows the respective photon flux. It is noticeable that the calibration factor of the monitor increases by about a factor of 3 between the two lowest photon energy values and the higher ones. This is accompanied by a drop in photon flux for the higher energies, which is also a factor of 3.

The main reason for the reduced calibration factor of the monitor is the measured photodiode current, which was by a factor of 4 higher at the two lower photon energies than the currents measured at the higher photon energies. This increased signal was obtained for measurements with and without attenuators and was accompanied by a similar increased count rate of Si KL₂L₃ Auger electrons measured simultaneously with the photodiode and photon monitor signal. Fig. S1.2 shows the variation of the ratio of the count rate at 1607 eV electron energy to the K-shell photoabsorption cross-section of silicon normalized to the photon flux derived from the photodiode signal. The error bars indicate the standard deviation.

These electrons have a range in silicon of about 340 nm (estimated from the ESTAR data base [6] using power-law extrapolation). The mean free path length of photons in silicon is between 200 μm and 400 μm [2] for the photon energies used in the experiments. Therefore, the thickness of the silicon surface layer contributing to the measured signal of electrons at a given energy varies insignificantly with photon energy. Therefore, the number of detected Auger electrons is expected to be proportional to the number of K-shell photoabsorption interactions in the silicon surface layer that contribute to the signal, and the ratio of the count rate to the K-shell photoabsorption cross-section is expected be constant, as can be seen in Fig. S1.2.

The slight difference, on the order of 1.5 %, between the lower and at higher photon energies is due to a significant background of photoelectrons and their progeny. This accounts for about a quarter of the signal in the Auger peak maximum for 11.9 keV photons (Fig. S1.3). Since this background should decrease as the energy of the photons, and hence the photoelectrons, increases, the ratio shown in Fig. S1.2 is expected to increase slightly with photon energy.

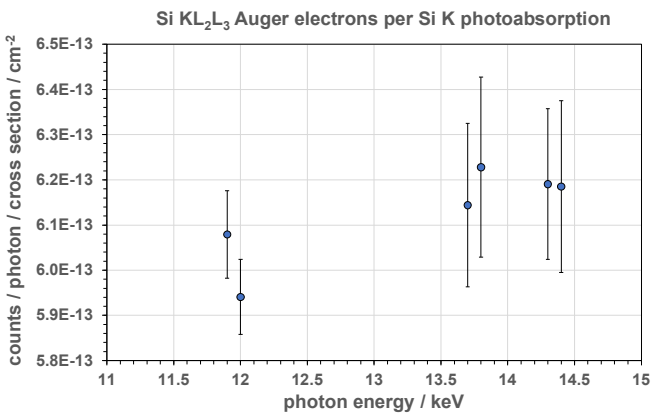

Fig. S1.2: Ratio of the number of detected electrons at 1607 eV (Si KL₂L₃) per incident photon to the K-shell absorption cross-section in the measurements at different photon energies for calibration of the photon flux monitor.

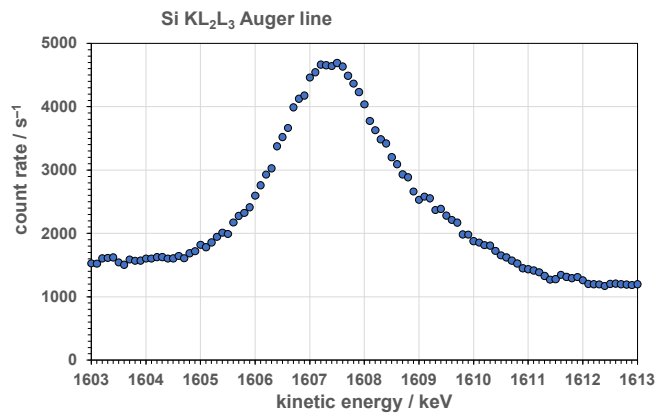

Fig. S1.3: Line profile of the silicon KL₂L₃ Auger line emitted from the silicon photodiode, measured with the HAXPES for a 11.9 keV photon beam at 45° grazing incidence.

The fraction of photons absorbed in the carbon foil of the photon flux monitor can be estimated from the linear attenuation coefficients in the NIST XCOM database [2] and varies from about 2.2 % at 14.4 keV to about 4.1 % at 11.9 keV. This means that the calibration factor of the monitor should be higher by a factor of about 2 at 11.9 keV than for 14.4 keV photons.

In fact, the ratio of the values of $Q_0$ given in Table S1.1 to the fraction of photoabsorbed photons is the same within a few percent for photon energies of 13.7 keV and higher, while values smaller by a factor of about 5 are obtained for this ratio at 11.9 keV and 12.0 keV.

The explanation finally found for these counterintuitive results was that the last focusing mirror was operated at a position where it was coated with boron carbide, whose reflectivity is lower than that of the intended palladium coating and decreases rapidly with increasing photon energy.

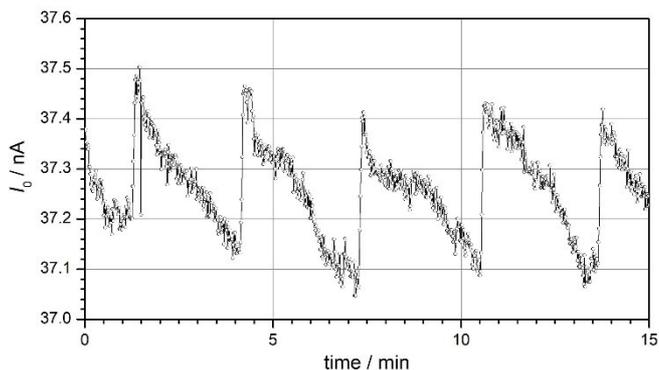

Fig. S1.4: Sample measurements with the photon flux monitor at 11.9 keV photon energy. The upward jumps occur when the beam current in the storage ring is topped up. The decrease between the jumps is due to electrons in the storage ring being scattered out of the stored beam by scattering on residual gas molecules.

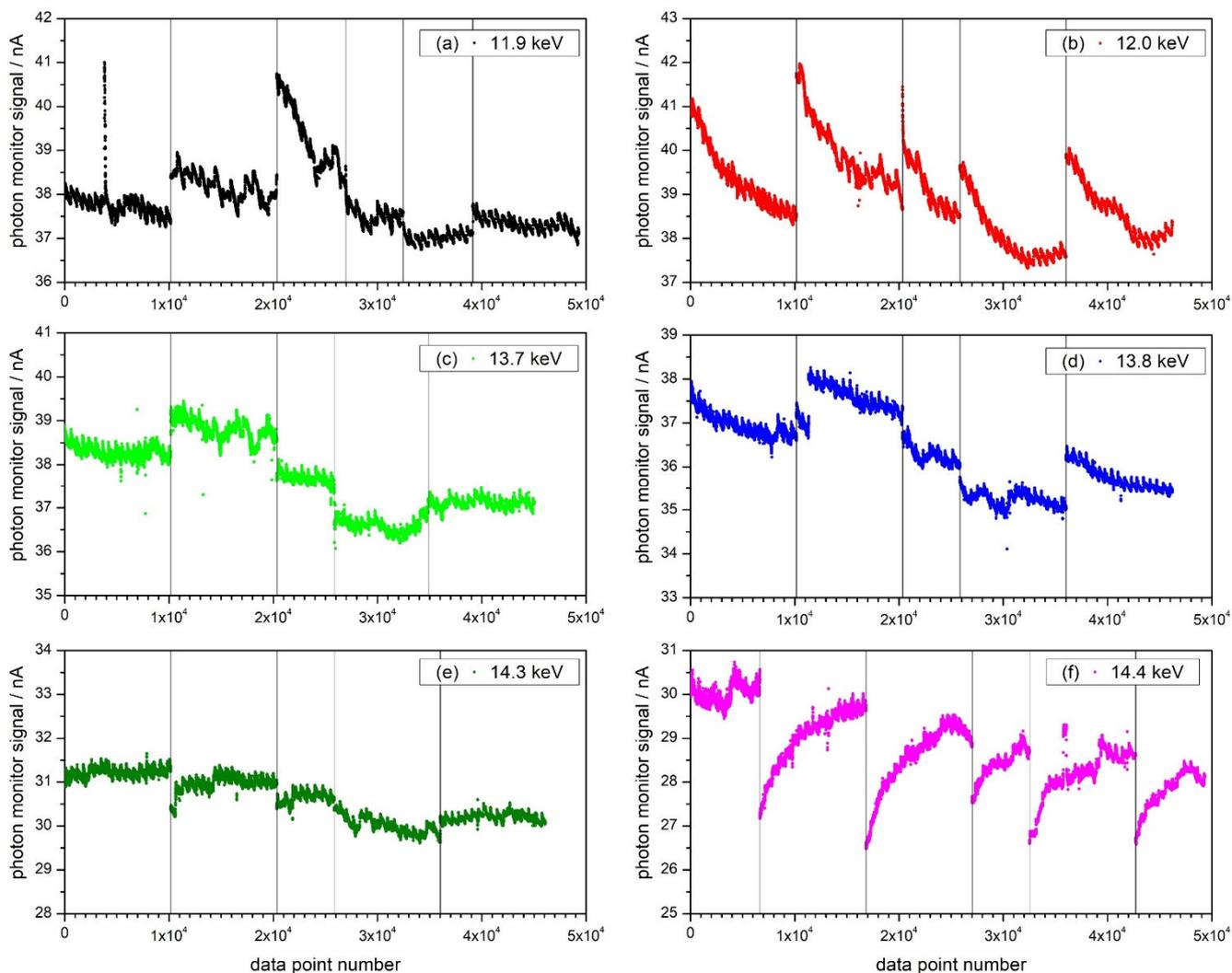

Fig. S1.5: Compilation of the photon flux monitor signal recorded during the measurements on the AuNP samples without beam attenuators at the six photon energies used in the experiments. The *x*-axis is in chronological order; the vertical lines delimit the sets of data points taken on particular positions or orientations of the sample(s).

This mirror is located downstream of the photon flux monitor, so that the photodiode at the measurement position in front of the electron spectrometer sees a lower photon flux than the photon flux monitor.

Sample measurements with the photon flux monitor are shown in Fig. S1.4 for a case where the photon beam was stable except for the continuous decrease due to the decrease of the electron current owing to scattering processes in the storage ring. The upward jumps are refills of the storage ring to maintain a constant average beam current of circulating electrons.

For a more comprehensive picture, Fig. S1.5 shows the recorded photon monitor signal values during all measurements on AuNP samples performed without beam attenuation. The vertical lines indicate the limits of different measurement runs, such as points in time when the sample under investigation, the position of the photon beam on the sample, its angle of incidence, or its energy were changed. The intervals for each photon energy are in chronological order, but the time elapsed between two measurement runs ending or starting at one of the vertical lines varies (between about six and 24 hours).

The strong drifts seen for the measurements at 12.0 keV and 14.4 keV and for one measurement run at 11.9 keV are related to a switching of photon energy, where the change of the undulator gap results in a different heat load on the monochromator crystals, which leads to transients until the actively controlled temperature settles back to the setpoint value. (The measurement sequence was generally started with 14.4 keV and ended at 11.9 keV.) Data collection after setting a new photon energy was delayed until the encoders of the monochromator and the undulator indicated their arrival at the desired positions. However, the drifts in the monitor signal indicate that the transients induced by the different heat load when the undulator gap changes, may take several tens of minutes to disappear. The results in Supplements 3 and 4 indicate that normalization of the signal with the photon monitor mitigates the effects of these beam distortions.

## Supplement 2: Determination of the photon beam size

To determine the size of the photon beam, auxiliary measurements were performed in which the photon beam was scanned horizontally and vertically across the edges of the silicon chip. In these measurements, the photocurrent of the diode and the photon flux monitor signal were recorded together with the count rate of the HAXPES, which was tuned to detect the most intense silicon KLL Auger line observed at 1607 eV ($KL_2L_3$ transition). For the horizontal scan, the photon beam was incident at a 5° grazing angle to the photodiode surface. For the vertical scan, an incidence angle of 30° was chosen. The different approach in the latter case was motivated by the presence of the electrical contacts at the respective chip edges.

The recorded position scans were analyzed using a linear regression of the measured photodiode signal in the region where the signal was rising between its two "plateaus". For the horizontal scan, the lower value was close to zero, and the value in the upper plateau, where the whole photon beam hit the silicon surface, was almost constant (Fig. S2.1 top). For the vertical scan, the lower value plateau corresponds to the photon beam hitting the detector on the aluminum contact outside the silicon chip. In this case, the signal in the plateaus had a significant slope (Fig. S2.1 bottom).

In the analysis, a linear regression was performed for the data points in the plateaus and for those data points between 10 % and 90 % of the difference between the plateaus (Fig. S2.1). The difference in position of the intersections between the lines fitting the rise and the two lines fitting the plateau regions for the horizontal and the vertical scan, $\Delta x$ and $\Delta y$, were then determined as

$$\Delta x = \frac{a_3 - a_2}{b_2 - b_3} - \frac{a_1 - a_2}{b_2 - b_1} \qquad (S2.4)$$

and analogously for $\Delta y$, where $a_n$ and $b_n$ are the intercept and the slope in the $n^{\text{th}}$ range of the linear regressions. The resulting values for $\Delta x$ and $\Delta y$ are $(103 \pm 5)$ μm and $(118 \pm 16)$ μm. The uncertainties have been determined by uncertainty propagation taking into account the correlation between the uncertainties of the parameters $a_n$ and $b_n$ of the same regression line.

It is assumed that the photon beam has an elliptical intensity profile given by a two-dimensional Gaussian distribution with standard deviations $\sigma_1$ and $\sigma_2$. Then, the effective beam area is given by

$$A = 2\pi\sigma_1\sigma_2 \qquad (S2.5)$$

In general, the principal axes of the ellipse may be tilted by an angle $\varphi$ with respect to the coordinate system. In the measurements, the marginal distributions are convoluted with a rectangular response function (proportion of the sensitive diode surface in the beam). These marginal distributions of intensity along the vertical and horizontal directions are also Gaussian distributions with standard deviations given by

$$\sigma_{x/y} = \sqrt{\frac{1}{2}\left(\sigma_1{}^2 + \sigma_2{}^2 \pm (\sigma_1{}^2 - \sigma_2{}^2)\cos 2\varphi\right)} \qquad (S2.6)$$

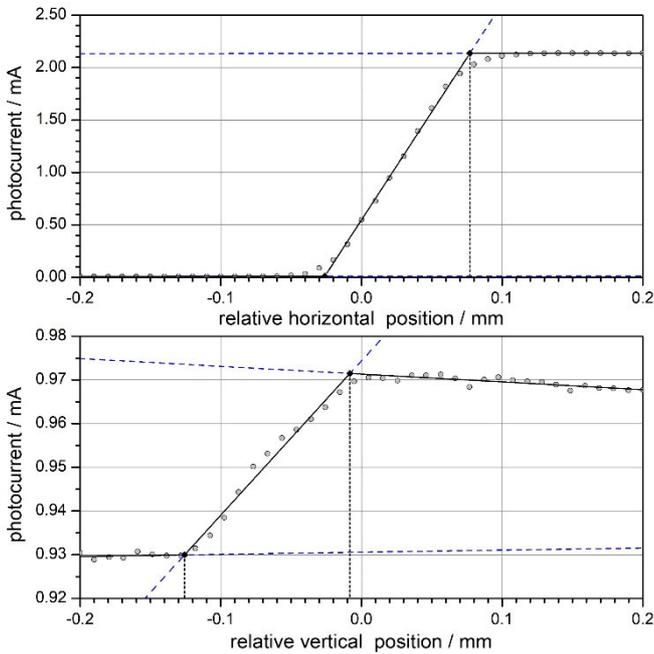

Fig. S2.1: Determination of the photon beam size from the variation of the signal when the photon beam was scanned across the edges of the photodiode's silicon chip. The results for the horizontal and vertical widths, $\Delta x$ and $\Delta y$, are $(103 \pm 5)$ μm and $(118 \pm 16)$ μm, respectively. The corresponding values of full width at half maximum for a Gaussian beam profile are $(97 \pm 5)$ μm and $(110 \pm 16)$ μm.

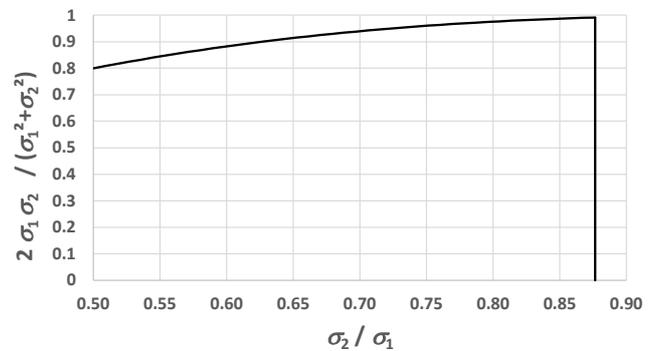

Fig. S2.2: Variation of the correction factor for the effective beam area in eq. (S2.9) with the ratio of the standard deviations along the principal axes of the beam profile, when this ratio varies between 0.5 and the maximum possible value $\Delta x/\Delta y$.

Therefore, the slope at the inflection points of the normalized cumulative distributions is $1/(\sqrt{2\pi}\sigma)$, where $\sigma$ is the standard deviation. In consequence, the distance between the intersection points is $\sqrt{2\pi}\sigma$ and

$$\Delta x = \sqrt{\pi(\sigma_1{}^2 + \sigma_2{}^2 + (\sigma_1{}^2 - \sigma_2{}^2)\cos 2\varphi)} \quad (S2.7)$$

$$\Delta y = \sqrt{\pi(\sigma_1{}^2 + \sigma_2{}^2 - (\sigma_1{}^2 - \sigma_2{}^2)\cos 2\varphi)} \quad (S2.8)$$

From these identities, the following expression for the effective beam area $A$ is obtained:

$$A = \frac{\Delta x^2 + \Delta y^2}{2} \frac{2\sigma_1\sigma_2}{\sigma_1{}^2 + \sigma_2{}^2} \quad (S2.9)$$

This allows us to estimate the variation of the true effective beam area with the ratio of the two standard deviations of the intensity distribution. Assuming $\sigma_1 > \sigma_2$, this variation is shown in Fig. S2.2 for ratios $\sigma_2/\sigma_1$ between 0.5 and the maximum possible value $\Delta x/\Delta y$. It is assumed that the potential values of the ratio are restricted to this interval with uniform probability density.

The resulting expectation and standard deviation of this correction factor is then obtained as $0.91 \pm 0.13$. With this correction factor, the final effective beam area is obtained from Eq. (S2.9) as $(1.11\pm0.22)\times10^{-2}$ mm².

## Supplement 3: Spectrometer transmission

The trajectory of an electron emitted from the sample into the solid angle accepted by the spectrometer is determined by its start position $r$, its initial energy $E$ and its direction of motion $\boldsymbol{\Omega}$ as well as the analyzer's pass energy $E_a$ and the retarding ratio $R$, which influences the focusing properties of the spectrometer's lens system. An electron is detected if its trajectory reaches the electron detector behind the hemispherical condenser without being intersected by the confining mechanical elements of the apertures involved, those being the slits, analyzer hemisphere electrodes, and mechanical parts of the retarding lens.

The spectrometer transmission function $t(r,\boldsymbol{\Omega},E|E_a,R)$ is unity if the trajectory reaches the detector in the aforementioned sense, and otherwise it is zero [1]. The measured count rate $\dot{N}_m$ for a given photon flux $\dot{N}_p$ and a set of retarding ratio $R$ and analyzer pass energy $E_a$ is given by

$$\dot{N}_m = \dot{N}_p \iiint t(r,\boldsymbol{\Omega},E|E_a,R) \frac{d^3\epsilon(r,\boldsymbol{\Omega},E)}{dE\,dA\,d\Omega} dA d\Omega\, dE \quad (S3.1)$$

where $\dot{N}_p d^3\epsilon(r,\boldsymbol{\Omega},E)/dEdAd\Omega$ is the number of emitted electrons per time, per energy interval $dE$ centered at $E$, per area element $dA$ located at position $r$ on the sample, and per solid angle $d\Omega$ along direction $\boldsymbol{\Omega}$. It is a function of the variables $E$, $r$, and $\boldsymbol{\Omega}$.

The relative variation $w(r,\boldsymbol{\Omega},E)$ of this function, defined as the ratio of $d^3\epsilon(r,\boldsymbol{\Omega},E)/dEdAd\Omega$ with respect to its maximum value $d^3\epsilon/dAdEd\Omega$, can be used to rewrite Eq. (S3.1) as follows:

$$\dot{N}_m = \dot{N}_p \frac{d^3\varepsilon}{dA\,dE\,d\Omega} T(E_a,R) \quad (S3.2)$$

with the (effective) spectrometer transmission $T(E_a,R)$. This quantity is given by

$$T(E_a,R) = \iiint t(r,\boldsymbol{\Omega},E|E_a,R) w(r,\boldsymbol{\Omega},E) dA\, d\Omega\, dE \quad (S3.3)$$

and has dimensions of emittance times energy. It depends on the functional dependence of $w$ on its arguments, that is, on the spectral-shape, the photon beam-shape and the angular distribution of the emitted electrons.

When the spectrometer is operated in the fixed retarding ratio mode, $T(E_a,R)$ is proportional to the analyzer pass energy $E_a$, so that the following identity holds:

$$T(E_a,R) = \frac{E_a}{E_{a,r}} T(E_{a,r},R) \quad (S3.4)$$

For the HAXPES spectrometer, the spectrometer's axis is aligned with the polarization vector of the linearly polarized photon beam. Therefore, only a minor variation of $w$ with the direction of emission is expected. (For a cos² dependence of the emission, the variation over the 9° angle of acceptance amounts to about 0.25 %.) On the contrary, with respect to the energy variation, there are two extreme cases of a sharp peak or a plateau, which lead to significantly different transmission values [2]. Since the spectral features observed in the spectra of this study have an energy width that is much larger than the spectrometer's energy resolution, the case of slowly varying continuous energy distributions appears more relevant. The spatial variation of $w$ can be assumed to roughly follow the intensity profile of the photon beam.

An evaluation of Eq. (S3.3), for example, by simulation, requires detailed knowledge of the interior geometry of the spectrometers, particularly in the lens system. In addition, the applied voltages on the different parts of the electron lens system are also needed. As the spectrometer manufacturer was not willing to disclose these technical details, the envisaged detailed determination of the spectrometer transmission could not be performed. Since the sample position at the experimental station is 8 mm closer to the spectrometer than the nominal working distance to enable a higher angular acceptance, the values given in the beamline design report [3] could also not be used. However, the manufacturer was kind enough to perform simulations of the spectrometer

Table S3.1: Calculated spectrometer transmission with a sample position that is 8 mm closer to the spectrometer aperture than the nominal working distance, for analyzer pass energy $E_a = 100$ eV and different retarding ratio $R_s$. (Courtesy of SPECS Inc.). The third column shows the transmission ratios for the reference retarding ratio $R_r = 50$ to the retarding ratio $R_s$ given in the first column obtained from measurements with different retarding ratios in overlapping energy ranges. The last column is the predicted transmission value for $E_a = 100$ eV and $R_r = 50$. The last line shows the mean and the standard deviation of the values in the fourth column. The value for $R_s = 50$ was used as a reference transmission value in the data analysis, and the ratio of the standard deviation to the mean value listed in the last line is taken as its relative uncertainty.

| $R_s$ | $T(E_a, R_s)$ / mm² sr eV | $G_{r,s}$ | $T(E_a, R_r)$ / mm² sr eV |
|---|---|---|---|
| 1 | $1.22\times10^{-4}$ | $1.11 \pm 0.05$ | $(1.35 \pm 0.06)\times10^{-4}$ |
| 2 | $1.87\times10^{-4}$ | $0.97 \pm 0.04$ | $(1.81 \pm 0.08)\times10^{-4}$ |
| 5 | $3.26\times10^{-4}$ | $0.64 \pm 0.03$ | $(2.10 \pm 0.08)\times10^{-4}$ |
| 10 | $3.97\times10^{-4}$ | $0.57 \pm 0.02$ | $(2.27 \pm 0.08)\times10^{-4}$ |
| 20 | $3.88\times10^{-4}$ | $0.65 \pm 0.02$ | $(2.52 \pm 0.08)\times10^{-4}$ |
| 50 | $3.34\times10^{-4}$ | $1$ | $3.34\times10^{-4}$ |
| 100 | $2.80\times10^{-4}$ | $1.44 \pm 0.07$ | $(4.02 \pm 0.20)\times10^{-4}$ |
| | | mean | $(2.49 \pm 0.85)\times10^{-4}$ |

transmission for one pass energy and a range of retarding ratios (Table S3.1).

In these simulations, a two-dimensional Gaussian source profile of 0.1 mm full width at half maximum was used along with a uniform variation of $w$ with energy and emission direction. In the experiments, the photon beam was incident at grazing angles of $\theta = 15°$ and $\theta = 60°$, respectively. The irradiated surface area is enlarged by $1/\sin\theta$. Since the photon beam dimensions (cf. Supplement 2) are very close to those used in the simulations, the transmission can be assumed to scale proportionally to the ratio of the irradiated areas in the experiment and the electron trajectory simulations.

$$T(E_a, R|A_b, \theta) = \frac{A_b}{\sin\theta}\frac{1}{A_r} \times T(E_a, R|A_r\pi) \qquad (S3.5)$$

When measurements are performed for the same electron energy with different retarding ratios $R_1$ and $R_2$, the ratio of the two (expected) count rates is obtained from Eqs. (S3.2) and (S3.4) as

$$\frac{\dot{N}_m(E|R_1)}{\dot{N}_m(E|R_2)} = \frac{R_2}{R_1}\frac{T(E_{a,r}, R_1)}{T(E_{a,r}, R_2)} \qquad (S3.6)$$

Defining $G(R_1, R_2) \equiv T(E_{a,r}, R_1)/T(E_{a,r}, R_2)$, this can be cast into

$$G(R_1, R_2) = \frac{R_1}{R_2}\frac{\dot{N}_m(E|R_1)}{\dot{N}_m(E|R_2)} \qquad (S3.7)$$

The values on the left-hand side of Eq. (S3.7) have been determined in the overlapping region of subsequent energy intervals, where spectra were measured with different retarding ratios. In addition, dedicated measurements were performed over larger energy ranges with the retarding ratios shown in the first column of Table S3.1. The resulting ratios $G(R_1, R_2)$ are shown in Fig. S3.2 as a function of the electron energy. From Fig. S3.2(a), it can be seen that for most of the combinations of retarding ratios, the ratio $G(R_1, R_2)$ is constant over the complete range of overlapping energies. The exception is the ratio between $R = 1$ and $R = 10$, where a decreasing trend is seen. This can be attributed to the measurement with $R = 1$ being the first in the series, where strong drifts were found in the monitor signal after setting the respective photon energy of 14.4 keV (Fig. S1.5).

In the ratio $G(50,100)$ shown in Fig. S3.2(b), there also seems to be a decreasing linear trend with increasing electron energy. The relative variation over the 6 keV range is about 10 % and, thus, of similar magnitude to the noise. It is presumably due to drifts in the photon beam position over the measurement time of more than one hours that results in a change of signal owing to the inhomogeneity of the surface coverage with AuNPs. The slope of the linear regression (blue line in Fig. S3.2(b)) is considered as an indication of the contribution of the photon beam position's instability to the uncertainty budget of the measurement.

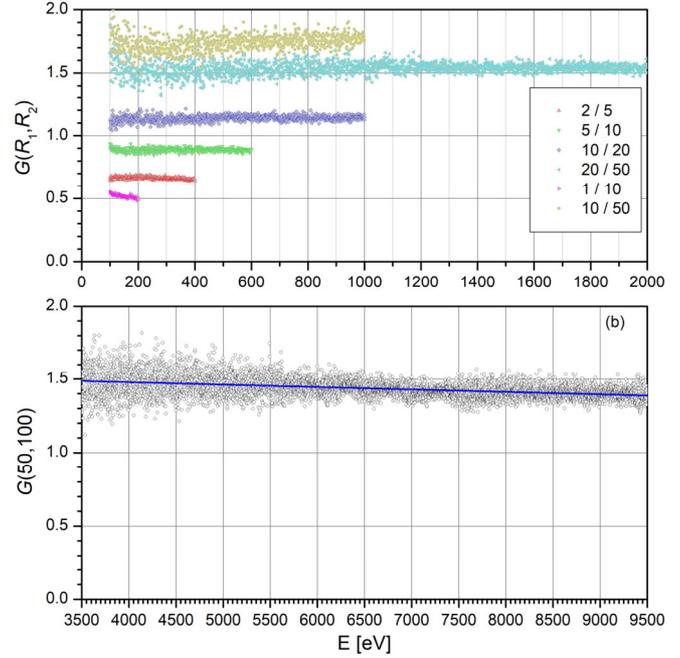

Fig. S3.1: Ratio of the measurement signal (normalized to the photon monitor signal and multiplied by the retarding ratio) for one retarding ratio to the signal for the second retarding ratio for different combinations.

The mean values of the datasets shown in Fig. S3.2 and their standard deviations were used to calculate the factor $G_{r,s}$ defined by

$$G_{r,s} \equiv G(R_r, R_s) \qquad (S3.8)$$

for a reference retarding ratio $R_r$ and the retarding ratios $R_s$ listed in the first column of Table S3.1. The respective values are listed in the third column of the table, and the last column shows the estimated transmission for 100 eV pass energy at the reference retarding ratio. The latter were calculated by multiplying the values in the second and third column, based on the relation

$$T(E_{a,r}, R_r) = G_{r,s} \times T(E_{a,r}, R_s) \qquad (S3.9)$$

which follows from the definition of $G_{r,s}$.

It can be seen from Table S3.1 that the values predicted for the transmission at the reference pass energy and retarding ratio show a large variation on the order of 20 % over the range of retarding ratios considered. This is particularly true of the values derived from the calculated transmittance at small retarding ratios, which are significantly smaller than those at higher retarding ratios. According to SPECs, this unexpected behavior for low retarding ratios is caused by the chromatic aberration of the hemispherical mirror mode. Since the simulations were performed for an idealized geometry that may be at variance with the actual one, the different values for the reference transmission were considered as equally likely to represent the true value. Therefore, their average was used as the best estimate for the reference transmission and their standard deviation was used as an uncertainty estimate according to the "Guide to the expression of uncertainties in measurement" (GUM) [4] The respective values that were used in the data analysis are shown in the last row of Table S3.1.

Fig. S3.2 shows the ratios $G(R_1, R_2)$ obtained by averaging over the overlap of two adjacent electron energy scans, that is, over the energy intervals between (a) 100 eV and 120 eV, (b) 1000 eV and 1200 eV, and (d) 6000 eV and 6500 eV. For the two energy scans with a retarding ratio of 50, there was only one common data point in this case. In this case (as shown in Fig. S3.2 (c)), the ratio was determined between the straight-line fits over the 200 eV intervals ending or starting at this energy and the measured data in the respective interval of the other energy range. The slope of the two lines was forced to be identical in the fit.

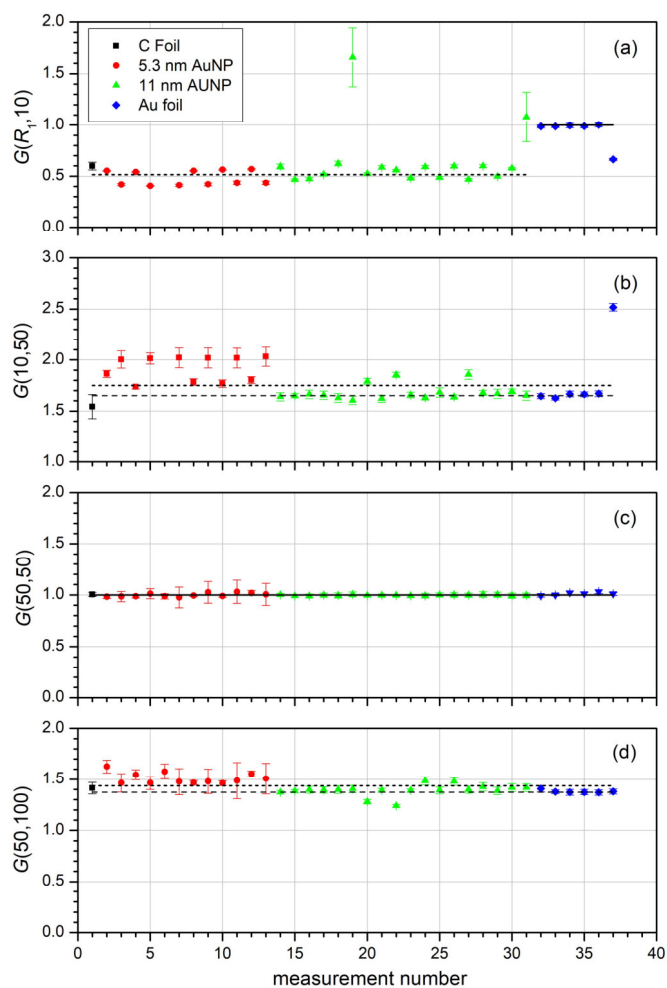

Fig. S3.2: Ratio of the signals measured with different retarding ratios in the energy intervals from (a) 100 eV to 120 eV, (b) 1000 eV to 1200 eV, (c) 3300 eV to 3700 eV, and (d) 6000 eV to 6500 eV. In (a) the retarding ratio $R_1$ is generally 1, except for the measurements with the gold foil (blue diamonds), where a retarding ratio of 10 was used for $R_1$. In (a), (b), and (d), the data points indicate the mean of the ratio of the signals multiplied with the respective retarding ratio (i.e. normalized to the analyzer pass energy), the uncertainty bars indicate the sample standard uncertainty. Since there was only one common data point in (c), linear extrapolations of the respective other energy range were used (see main text for details).

The short-dashed lines in Fig. S3.2(a), (b), and (d) indicate the mean values obtained from the data shown in Fig. S3.1. The long-dashed lines in (b) and (d) indicate the median of the green and blue data points, which deviate by about 6 % and 5 % from the former values.

The solid lines in Fig. S3.2(a) and (c) indicate unity. This is the expected value since, in these measurements, the retarding ratio in the first and second energy range were identical. It is seen that generally the ratios agree well with the expected value within the statistical uncertainty with the exception of the 14.4 keV measurement on the gold foil, where a value of about 0.66 is found. This outlier is accompanied by a corresponding outlier in Fig. S3.2(b), which is higher than the values found for other energies and samples by a factor of about 1.6.

The origin of these aberrant transmission ratios is related to an abnormal time evolution of the photon monitor signal, which is depicted in Fig. S3.3. The photon monitor signal was significantly lower in these measurements because attenuators were used in the photon beam to prevent an overloading of the HAXPES detector. The vertical jumps are due to topping-up the electron current in the storage ring. It can be seen that for the first approx. 2500 data points, the measured signal is almost constant between subsequent steps, and that the value in the plateaus increases with time. For the later data points between about 2600 and 3400, the typical decrease due to residual gas scattering is seen. This initial increase in the monitor signal is attributed to transients in the trajectories of the electron beam after changing the undulator gap to match the photon energy set on the monochromator, which was seen in Fig. S1.5.

While the transmission ratios shown in Fig. S1.5 for the AuNP samples are generally in reasonable agreement with the values obtained from Fig. S3.1, two major outliers are seen in Fig. S3.2 (a) for the 11 nm AuNP sample. These data points are also for measurements at 14.4 keV (for 60° grazing incidence). In fact, these measurement runs correspond to the monitor signals in the second and third intervals in Fig. S1.5(f), where there was a particularly large drift of the photon monitor current and ensuing large changes in the electron beam emittance.

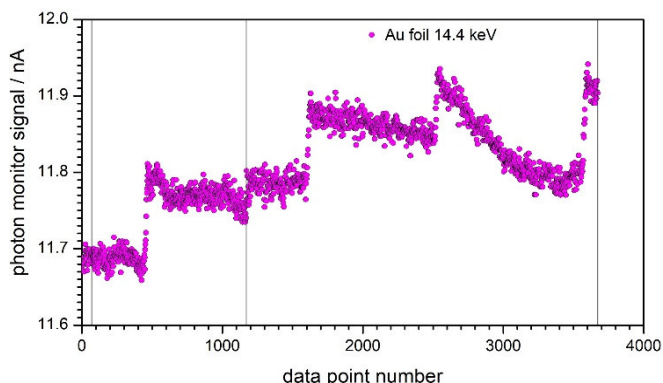

Fig. S3.3: Recorded photon monitor signal during the measurements with the gold foil at 14.4 keV photon energy. The vertical lines indicate the start of a new energy scan over the intervals given in Table 1.

Smaller but significant deviations in the order of tens of percent are seen in Fig. S3.2 (a) and (b) for the 5.3 nm AuNP sample, where the measurements at 60° grazing incidence are systematically lower in (a) and higher in (b). There are also deviations for some of the measurements with the 11 nm AuNP sample. These deviant values are attributed to different transmission values for the electrons of the background signal for which the spatial extension of the source distribution cannot be assumed to be similar to the photon beam profile.

All these deviations are considered as uncertainties of the spectrometer transmission that apply to the individual measurement on top of the global uncertainty from the determination of the transmission, which is given in Table S3.1.

## Supplement 4: Sample non-uniformity

Fig. S4.1 shows the data measured during position scans of the 11 nm AuNP sample, where the photon beam was incident at a grazing angle of 15° to a sample surface.

Fig. S4.1(a) and (b) show measurements at the Au L3 photoelectron line taken before the first electron energy spectra were recorded. The sample was scanned along the horizontal ($x$) and vertical ($y$) directions. These scans were performed to identify a position on the sample yielding a high signal. The respective measurement position indicated by the dot-dashed lines corresponds to the data labelled Pos.1 in Fig. 7 and Fig. 8. It should be noted that the measurements in (b) preceded those in (a).

Fig. S4.1(c) shows data measured while the sample was scanned in the direction of the photon beam such as to optimize its position with respect to the electron energy analyzer. This measurement was taken after a longer period of beam loss and was followed by the vertical scan shown in Fig. S4.1(d). The respective measurement position indicated by the dot-dashed lines corresponds to the data labelled Pos.3 in Fig. 7(a) and is in the center of the surface area covered with AuNPs.

Fig. S4.1(e) and (f) correspond to measurements at the energy of the Au L3M5M5 Auger line (7447 eV), where it was ignored that this line is not produced by 11.9 keV photons. The detected electrons are therefore from the low-energy tail

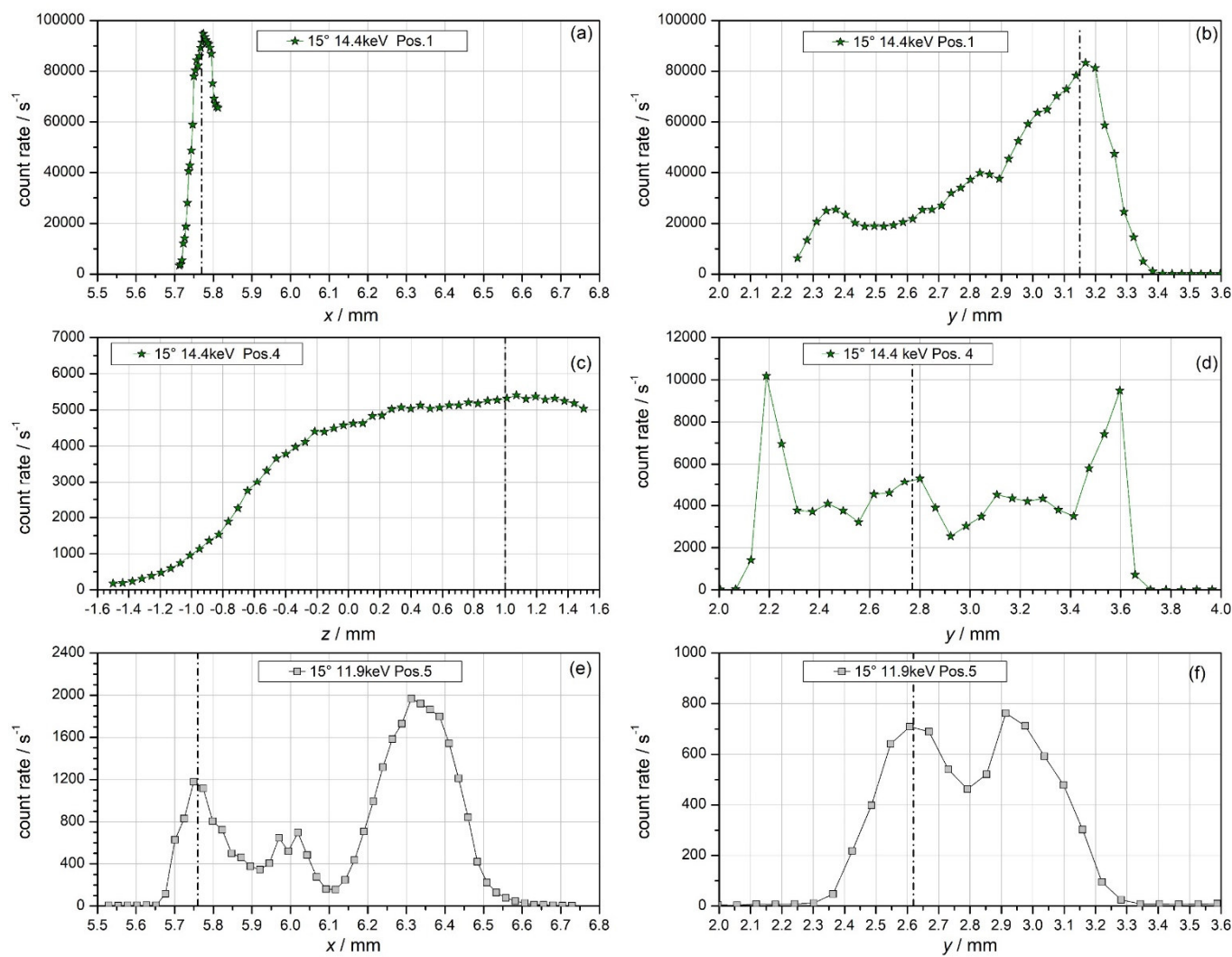

Fig. S4.1: Signal measured during position scans with the 11 nm AuNP sample in the horizontal ($x$) and vertical ($y$) directions perpendicular to the photon beam. The photon beam was incident at a grazing angle of 15° to the sample surface. Graphs (a) and (b) show measurements at the Au L3 photoelectron line taken to identify a position on the sample yielding a high signal. Graph (c) shows the the Au L3 photoelectron signal measured while the sample was scanned in the direction of the photon beam such as to optimize its position with respect to the electron energy analyzer. The ensuing vertical is shown in (d). Graphs (e) and (f) show measurements at the energy of the Au $L_3M_5M_5$ Auger line (7447 eV). The vertical dot-dashed lines indicate the measurement positions chosen based on these position scans.

of the Au M-shell photoelectrons (Fig. S4.2). The respective measurement position indicated by the vertical dot-dashed lines corresponds to the data labelled Pos.5 in Fig. 7 and Fig. 8. Also in this case, the measurements shown in Fig. S4.1(f) preceded those shown in Fig. S4.1(e).

Fig. S4.3 shows the ratios of the effective particle radiance (obtained from the measurements at one position) to that at another position (as specified on the *y*-axis labels). The two measurements at 12.0 keV (Fig. S4.3(b)) gave almost identical results for energies above 2.5 keV, while larger discrepancies are seen below 0.5 keV and in the region in between the two energy values. The latter observation is explained by the fact that the measurement in the second energy range of Table 1 initially failed and had to be repeated at a later time with a slightly different measurement position. For the higher photon energies (Fig. S4.3(c) to (f)), the data for the higher photon energies appear to be shifted by an almost constant value on the logarithmic *y*-axis.

For the data measured at 11.9 keV photon energy, a discrepancy is observed in Fig. S4.3(a), which cannot be corrected by applying a constant factor. This discrepancy is presumably due to a different background contribution from electrons originating from photoabsorption in surface contaminants [28] or in the aluminum support of the samples.

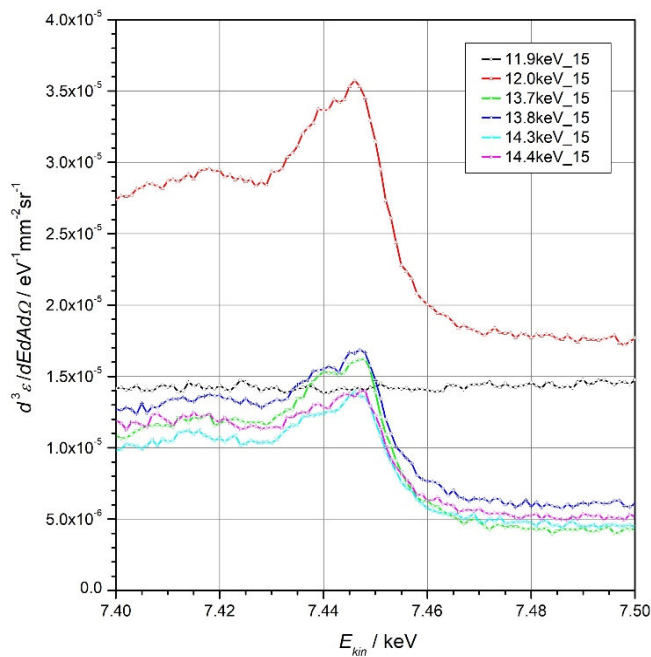

Fig. S4.2: Measured signal of the 11 nm AuNP sample (at Pos.2) around the experimentally observed energy position of the Au $L_3M_5M_5$ Auger line (Table 2) for the six photon energies used in the experiments.

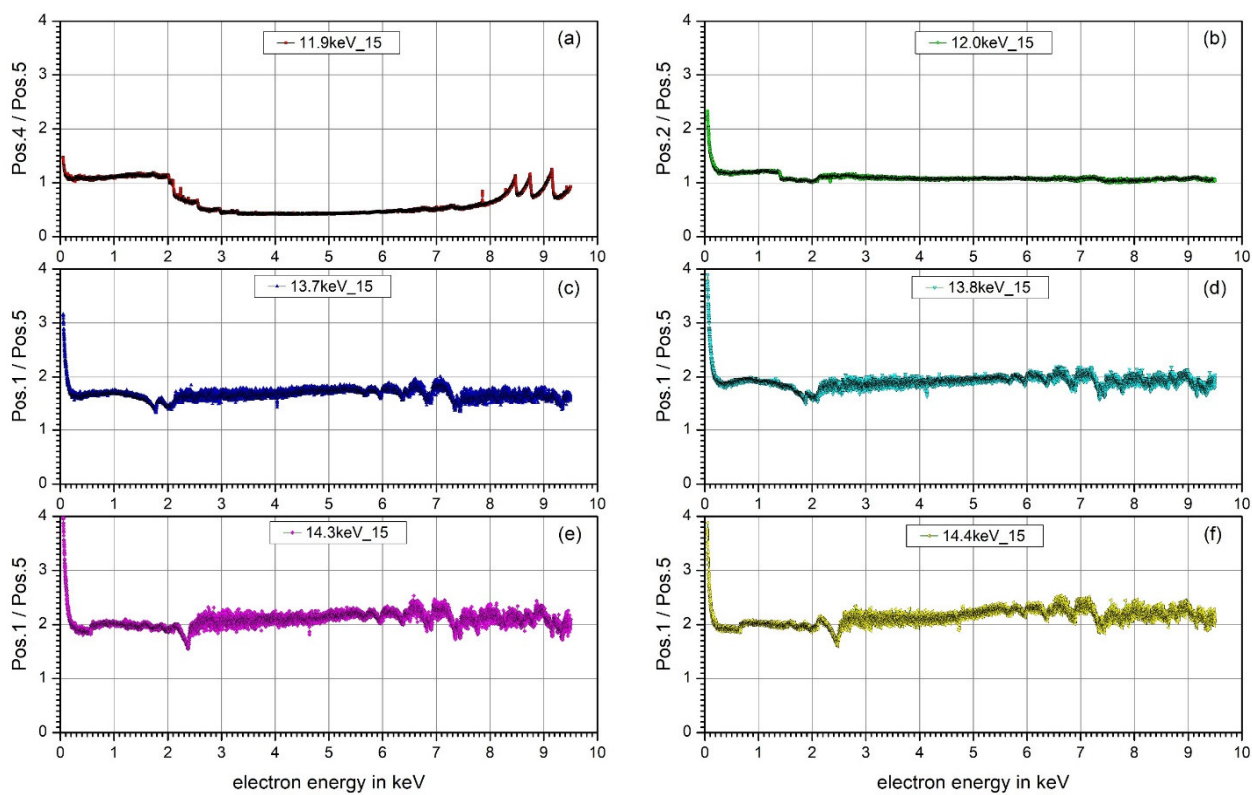

Fig. S4.3: Ratio of the spectral particle radiance of the 11 nm AuNP sample from measurement runs at 15° grazing angle of photon incidence. The data shown in (a) are for a measurement position in the plateau region of the AuNP coverage, where monolayers were found in STEM images and for a position closer to the outer rim of the region covered with AuNPs, where a larger proportion of multilayers was found. The results shown in (b) are for two measurments at about the same position in region covered with monolayers of AuNPs and suggest that, for energies above a few hundred eV, data are repeatable within the variation of AuNP surface coverage. Panels (c) to (f) are for the previously mentioned position with monolayer coverage and a position of maximum signal, where multilayers are prevalent. The different density of AuNPs appear to essentially increase the values by a constant factor for energies above a few hundred eV.

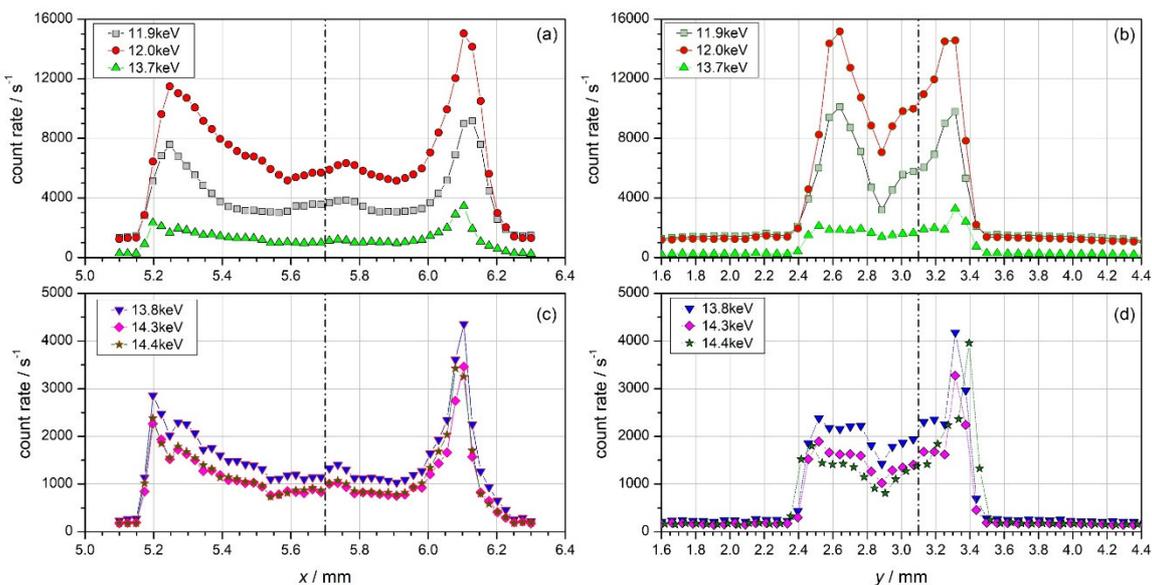

Fig. S4.4: Signal measured during position scans with the 11 nm AuNP sample in the horizontal ($x$) and vertical ($y$) directions perpendicular to the photon beam. The photon beam was incident at a grazing angle of 60° to the sample surface, and the detected electron energy was 7447 eV, which corresponds to the Au $L_3M_5M_5$ Auger line (Table 2). Based on these measurements, the position indicated by the dot-dashed lines was chosen for recording the electron energy spectra.

The position scans for finding the measurement position in the 60° grazing incidence measurements are shown in Fig. S4.4). In this case, the scans were performed at all investigated photon energies to check for potentially different beam positions on the sample when the monochromator was tuned to a different energy. Fig. S4.4(d) shows a vertical-shift of the 14.4 keV photon beam, while for all other energies and the horizontal direction, no apparent shift can be seen. It must be noted, however, that the step size was about 25 µm in the horizontal and about 60 µm in the vertical directions, while the beam size was about 100 µm (Supplement 2).

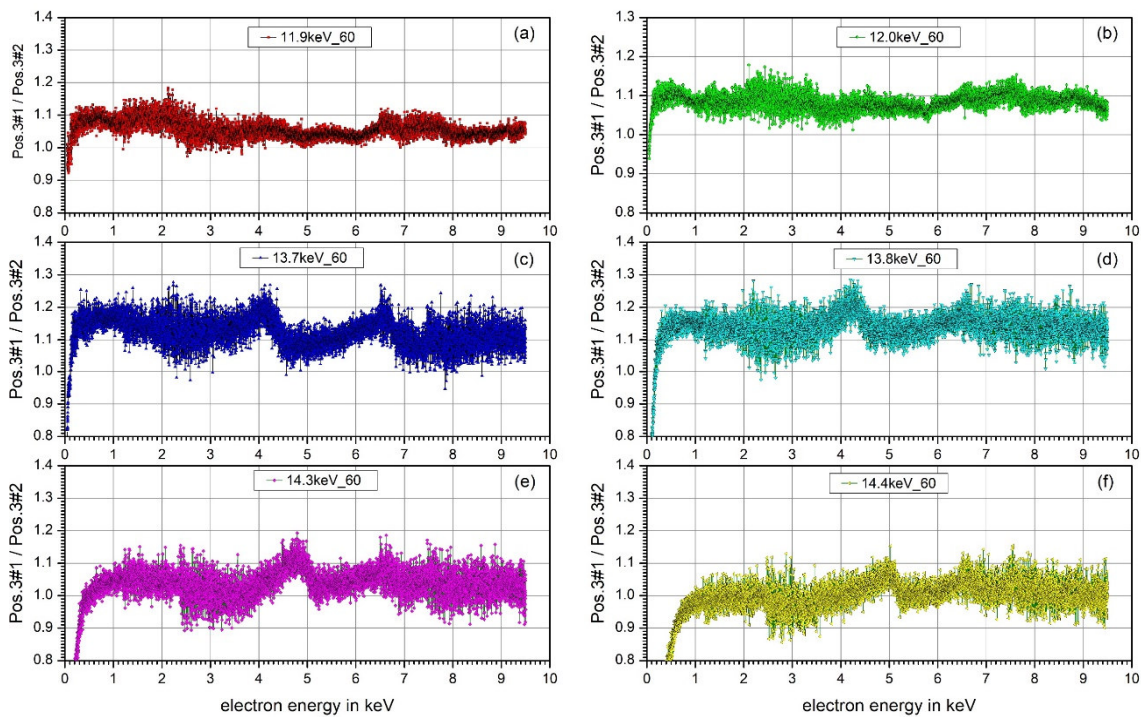

Fig. S4.5: Ratio of the spectral particle radiance of the same position on the 11 nm AuNP sample from two subsequent measurement runs at a 60° grazing angle of photon incidence. The data show that (for energies above a few hundred eV) results are repeatable within the variation of AuNP surface coverage which is constant in the central region of the sample within ± 15 %.

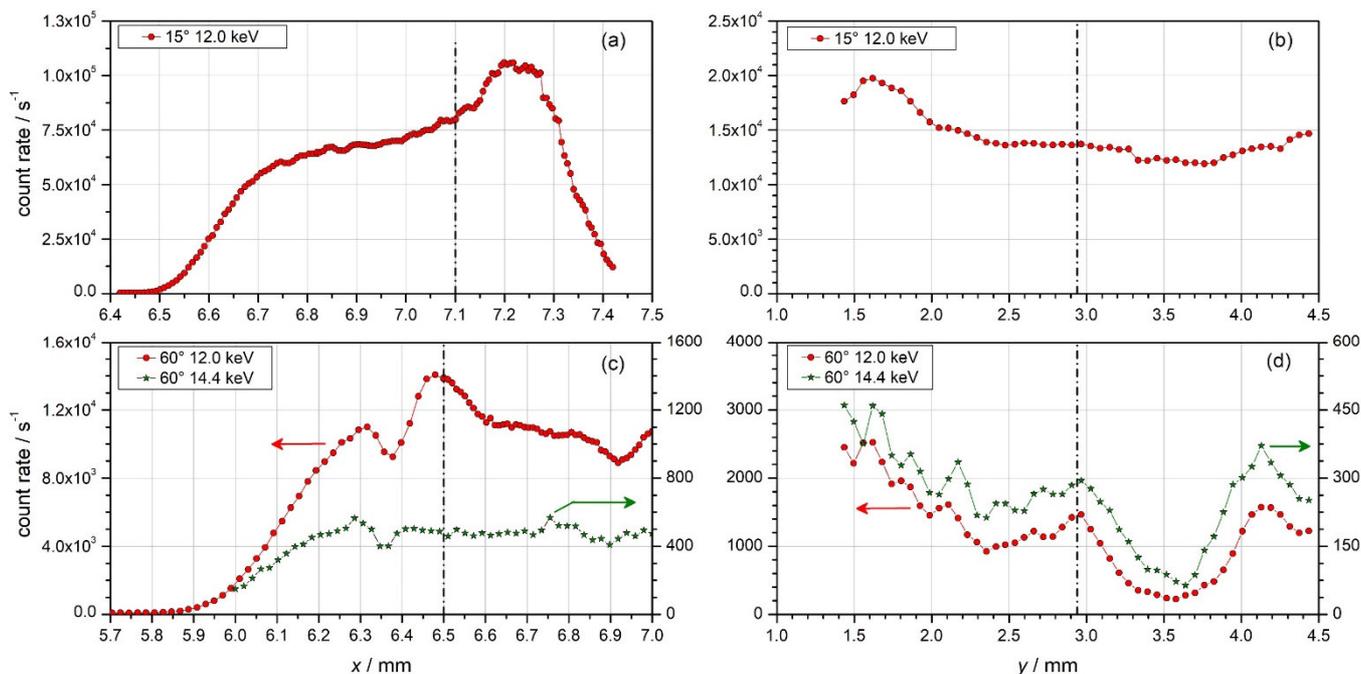

Fig. S4.6: Signal measured during position scans with the 5.3 nm AuNP sample in the horizontal ($x$) and vertical ($y$) directions perpendicular to the photon beam. Panels (a) and (b) show measurements with a 12.0 keV photon beam incident at a grazing angle of 15° to the sample surface, while the grazing incidence angle was 60° for panels (c) and (d) for photon beams of energy 12.0 keV and 14.4 keV, respectively. The detected electron was that of the Au $L_3$ photoelectron line energy for the data in (a) and (c), whereas for the vertical scans shown in (b) and (d), the Au $L_3M_5M_5$ Auger line (7447 eV) was detected.

The ratios of the effective particle radiance obtained from the measurements at nominally the same position in the first run to that in the second run are presented in Fig. S4.5. It can be seen that except for the lowest energies the ratios are close to unity with relative deviations less than 15 %. The larger extension of the electron energy range, where larger discrepancies are seen in Fig. S4.5(f), appears related to the strong drift in photon flux that can be seen in Fig. S1.5(f). This also caused the outliers found in Fig. S3.2. The structures that can be seen between 4 keV and 7 keV in Fig. S4.5(c) to (f) are related to the Tougaard background of Cu K-shell photoelectrons.

Position scans on the sample with 5.3 nm AuNPs are shown in Fig. S4.6. The data from the measurements at a 15° grazing incidence angle shown in Fig. S4.6(a) and (b) vary within a few tens of percent in the region where a high signal was detected for the Au L3M5M5 Auger line. The measurements at 60° grazing incidence (Fig. S4.6(a) and (b)), where the irradiated area is much smaller than for 15° incidence, however, also show inhomogeneities of almost an order of magnitude. This was also found in the extreme case shown for the 11 nm sample in Fig. S4.1(e).